\documentclass[final,1p,times]{elsarticle}

\usepackage{graphicx}
\usepackage{amssymb}

\usepackage{lineno}

\usepackage{hyperref}

\biboptions{sort&compress,numbers}

\usepackage{appendix}
\usepackage{bm}
\usepackage{amsmath}
\usepackage{amsfonts,accents}
\usepackage{mathrsfs}
\usepackage{comment}
\usepackage{color}
\usepackage{jabbrv}
\usepackage{hyperref}
\hypersetup{
    colorlinks=true,
    linkcolor=blue,
    filecolor=magenta,      
    urlcolor=cyan,
}

\newcommand{\mc}[1]{{\mathcal #1}}

\newcommand{\mr}{\mathring}

\journal{
Journal of the Mechanics and Physics of Solids}

\begin{document}

\begin{frontmatter}


\title{A Kirchhoff-like theory for hard magnetic rods under geometrically nonlinear deformation in three dimensions}



\author[flexlab]{Tomohiko G. Sano\footnote{Current affiliation: Laboratory for Structural Instabilities in Geometry and Materials, Department of Mechanical Engineering, Keio University, 3-14-1 Hiyoshi, Kohoku-ku, Yokohama, Kanagawa, 223-8522, Japan}}
\author[flexlab]{Matteo Pezzulla\footnote{Current affiliation: Slender Structures Lab,
	Department of Mechanical and Production Engineering,
	{\AA}rhus University,
	Inge Lehmanns Gade 10, 8000 {\AA}rhus C, Denmark}}
\author[flexlab]{Pedro M. Reis}
\ead{pedro.reis@epfl.ch}

\address[flexlab]{Flexible Structures Laboratory, Institute of Mechanical Engineering, 
\'Ecole polytechnique f\'ed\'erale de Lausanne, Switzerland.}

\begin{abstract}
Magneto-rheological elastomers (MREs) are functional materials that can be actuated by applying an external magnetic field. 
MREs comprise a composite of hard magnetic particles dispersed into a nonmagnetic elastomeric (soft) matrix. By applying a strong magnetic field, one can magnetize the structure to program its deformation under the subsequent application of an external field. There is a variety of types of MREs depending on the value of their coercivity (\emph{i.e.} the necessary field strength to erase the magnetization) that can be broadly classified into soft or hard MREs. 
{\it Hard} MREs, whose coercivities are large, have been receiving particular attention because the programmed magnetization remains unchanged upon actuation. 
Hence, once a structure made of a hard MRE is magnetized, it can be regarded as magnetized permanently.
Motivated by a new realm of applications, there have been significant theoretical developments in the continuum (3D) description of hard MREs. 
By reducing the 3D description into 1D or 2D via dimensional reduction, several theories of hard magnetic slender structures such as linear beams, elastica, and shells have been recently proposed. 
In this paper, we derive an effective theory for MRE rods (slender, mono-dimensional structures) under geometrically nonlinear 3D deformation. Our theory is based on reducing the 3D magneto-elastic energy functional for the hard MREs into a 1D Kirchhoff-like description (centerline-based). 
Restricting the theory to 2D, we reproduce previous works on planar deformations. 
For further validation in the general case of 3D deformation, we perform precision experiments with both naturally straight and curved rods under either constant or constant-gradient magnetic fields. Our theoretical predictions are in excellent agreement with both discrete simulations and precision-model experiments. 
Finally, we discuss some limitations of our framework, as highlighted by the experiments, where long-range dipole-dippole interactions, which are neglected in the theory, can play a role. 
\end{abstract}


\end{frontmatter}


\newpage

\section{Introduction}

Leveraging the coupling between internal degrees of freedom and material strains to open up new avenues for structural instabilities has been receiving much recent attention in the mechanics and advanced functional material communities \cite{Li2009IEEE,Dorfmann:2014}. The magneto-elastic coupling is particularly interesting since it can be exploited to actuate soft magnetic structures without any mechanical contact. Magneto-rheological elastomers (MREs) comprise a composite of magnetized (metal) particles and a soft elastic matrix that can respond to an applied magnetic field. 
MREs have long been studied, from both mathematical modeling and application perspectives~\cite{Dorfmann:2014,Menzel:2019iu}.
For instance, tunable mechanical functionalities of magnetic structures have been used in the applications such as sensing medical devices~\cite{Li2009IEEE}, engine mounts~\cite{Ginder:1999ku}, soft robotics~\cite{Huang:2016ee}, actuators~\cite{Hu:2018fe}, and minimally invasive procedures~\cite{Pancaldi:2020gt}.

The mechanical responses of MREs to an applied field can be categorized into the following (i)-(v) cases, depending on their ingredients, such as the coercivity, which is the necessary field strength to erase the magnetization. 
(i) {\it Super-para-magnetic} and {\it soft-ferromagnetic materials} have large magnetic susceptibility, enabling the temporary magnetization by an external magnetic field~\cite{Moon:1968da, Cebers:2003en, CebersPRE2004, Dreyfus:2005hc, Roper:2006fp,  CebersPRE2007, Gerbal:2015ht}; 
a super-para-magnetic material loses its magnetization, once the external field is removed. As a result, the magnetization of these materials depends strongly on the external fields. 
(ii) {\it Hard-ferromagnetic materials} have a high coercivity. An external field does not change the magnetization within the range of practical actuation. Hence, their magnetization is independent of external fields. 
(iii) {\it  Chains of hard-ferromagnetic spheres} are purely magnetic systems without elasticity~\cite{Hall:2013gm,Vella:2014ix,Schonke:2017if}. The complex internal interaction among beads yields a variety of equilibrium configurations. Interestingly, these chains posses an effective bending stiffness \cite{Vella:2014ix}.
(iv) {\it MREs embedded with soft-ferromagnetic
particles} or {\it soft magnetic elastomer}~\cite{Rigbi:1983hu,Ginder:1999ku,Dorfmann:2003gm,Li2009IEEE,Danas:2012hs,Loukaides:2015im,Seffen:2016dl,Schmauch:2017dd,Ciambella:2018bs,Singh:2018dq,Psarra:2019jp,Alapan:2020fm} deform upon the application of an external magnetic field due to elasto-magnetic couplings. 
The magnetic particles with low coercivity dispersed in a soft matrix align themselves with the applied field, thereby forming particle chains. To enhance the elasto-magnetic coupling, it is common to apply the magnetic field to MRE during the curing processes.
(v) {\it MREs embedded with hard-ferromagnetic
particles} or {\it hard magnetic elastomer} have both high coercivity and flexibility in shape-programming ~\cite{Lum:2016fc,Kim:2018bu,Zhao:2019hk,Wang:2020du,Ciambella:2020je,Chen:2021cx,Yan:2021jr,Yan:2020prep,Pancaldi:2020gt}. 
Once they are saturated magnetically, they retain permanent magnetization. For the remainder of this paper, we will focus on this latter type (v).

The high coercivity of hard MREs enables the design of highly functional mechanical systems, such as auxetic metamaterials~\cite{Kim:2018bu}, programmable materials~\cite{Chen:2021cx}, micro-swimmers~\cite{Diller:2014fd,Hu:2018jq,Zhang:2018eh}, micro or soft-robotics~\cite{Huang:2016ee,Tsumori:2015ke,Hu:2018fe} and haptic devices~\cite{Pece:2017ib}.
Motivated by these emerging applications of hard MREs, it is timely to formulate structural theories for elementary building blocks made of hard MREs to serve as predictive tools in the design process and aid in subsequent analysis.

During the past decade, there have been significant developments in the fundamental theory of hard magnetic elastomers. 
In a recent pioneering work by Zhao et al.~\cite{Zhao:2019hk}, a nonlinear theory for finite deformation of 3D (bulk) hard MRE was developed, based on the nonlinear elasticity complemented by the magneto-mechanical constitutive laws. In this framework, the Helmholtz free energy density function comprises elastic (neo-Hookean) and magneto-elastic parts. The relationship between the induced magnetic flux density in the material and the external field strength is assumed to be linear, with a permeability close to that of vacuum. The free energy density for the magneto-elastic part is modeled such that the magnetic moment is embedded in a soft matrix. A simulation framework using the finite element method developed by the same authors, based on nonlinear elasticity in 3D, was shown to be in quantitative agreement with their experimental results~\cite{Zhao:2019hk}.

Based on the framework reported in Ref.~\cite{Zhao:2019hk}, theories for hard magnetic linear beams and elastica have been derived and validated against experiments under either constant magnetic field~\cite{Wang:2020du} or a field with constant gradient~\cite{Yan:2020prep}. 
Furthermore, it was shown how a hard magnetic beam (or elastica) can buckle under an applied field, analogous to an elastica under compression~\cite{Wang:1986kl,Lum:2016fc,Ciambella:2020je,Wang:2020du,Yan:2020prep}.
Methodologically, the derivations of all of the above reduced models for hard magnetic linear beam and elastica always boils down to the formal procedure of dimensional reduction~\cite{Dill:1992cp,Audoly:2010Book}, which is summarized next. 
More specifically, the total energy functional of a beam is modeled as Ref.~\cite{Zhao:2019hk}, with elastic and magneto-elastic parts. The integrand is expanded with respect to the thickness of the cross-section, only retaining the leading order terms. Upon integration over the cross-section, a one-dimensional (1D)-reduced energy is derived. 
The reduced framework is consistent with the analogous procedure based on force and moment balance~\cite{Lum:2016fc}, which was also implemented in the study of the inverse problem~\cite{Ciambella:2020je}. 
A similar strategy based on dimensional reduction has also been employed to predict the axisymmetric deformation of pressurized hard magnetic shells~\cite{Yan:2021jr}. 

While theoretical frameworks for one-dimensional hard magnetic slender structures are now well established for planar 2D deformations (e.g., beams and elastica), the modeling of hard magnetic rods, with natural curvatures, undergoing 3D deformation is still lacking. However, the extension of the formulation in 2D toward 3D deformation is not straightforward because of the complex interplay between elasticity and geometry~\cite{Dill:1992cp,Audoly:2010Book,Landau:1980,Powers:2010cv,Bigoni:2015Book,Goriely:2017Book}. The precise modeling of a {\it hard magnetic elastomeric rod} is crucial to simulate, for example, micro-magnetic-swimmers~\cite{Diller:2014fd,Hu:2018jq,Zhang:2018eh} in a 3D complex channel, and endovascular probes \cite{Pancaldi:2020gt}.

In this paper, we establish a theory for the deformation of {\it hard magnetic elastomeric rods}, combining dimensional reduction and simulations, both of which are validated against precision experiments. For our theory, we start from the total energy functional comprising the sum of the Hookean elastic energy~\cite{Audoly:2010Book} and the elasto-magnetic free-energy proposed in Ref.~\cite{Zhao:2019hk}. We then employ the centerline-based kinematics utilizing the Cosserat frame and Darboux vector~\cite{Goriely:2017Book}. Along the (circular) cross-section, we span the (local) polar coordinate around the centerline position to parametrize the material points of the hard magnetic rod. Subsequently, we expand the integrand of the elasto-magnetic energy for the rod diameter, retaining only the leading order terms. After integrating the integrand over the cross-section, we obtain the reduced elasto-magnetic functional. Based on this reduced elasto-magnetic functional, we then apply the principle of virtual work to arrive at the governing nonlinear ordinary differential equations (strong form of the problem). The derived equilibrium equations are analogous to the Kirchhoff rod equations, but with additional elasto-magnetic terms; hence, we call them the {\it magnetic Kirchhoff rod equations}. 
The nonlinear deformations based on the magnetic Kirchhoff rod equations are computed with the discrete simulation method~\cite{Chirico:1994hm,Wada:2007dt,Wada:2007kw,Vogel:2010gc,Reichert:2006Thesis,Bergou:2008dt,Morigaki:2016dp,Sano:2017ki,Sano:2018ji,Sano:2019hk}. Both the dimensional reduction procedure and the simulations are then validated against precision experiments. 
Since our theory is based on the assumption that the magnetic torque is induced by the misalignment between the magnetization and the applied field, the long-range dipole interactions are not modeled. 
Still, we find excellent agreement between theory and experiments as long as the long-range dipole interactions are negligible. We quantify when and how the long-range interactions become important in the experiments.

Our paper is organized as follows. In Sec.~\ref{sec:DefProb}, we define the problem. The derivation of the magnetic Kirchhoff equations follows in Sec.~\ref{sec:Theory}, where we also show that our theory reproduces the previous results of planar deformation of hard magnetic elastica and beams. To compute 3D geometrically nonlinear deformation, we numerically solve the derived set of equations as detailed in Sec.~\ref{sec:Sim}. In Sec.~\ref{sec:Exp}, the experimental fabrication, apparatus, and protocols are presented. We report the comparison between theory and experiment in Sec.~\ref{sec:3D}. We further discuss the limitation of our theory systematically in Sec.~\ref{sec:Limitation}. In Sec.~\ref{sec:Concl}, we discuss and summarize our findings. 

\section{Definition of the problem}
\label{sec:DefProb}

\begin{figure}[!ht]
    \centering
    \includegraphics[width=0.8\textwidth]{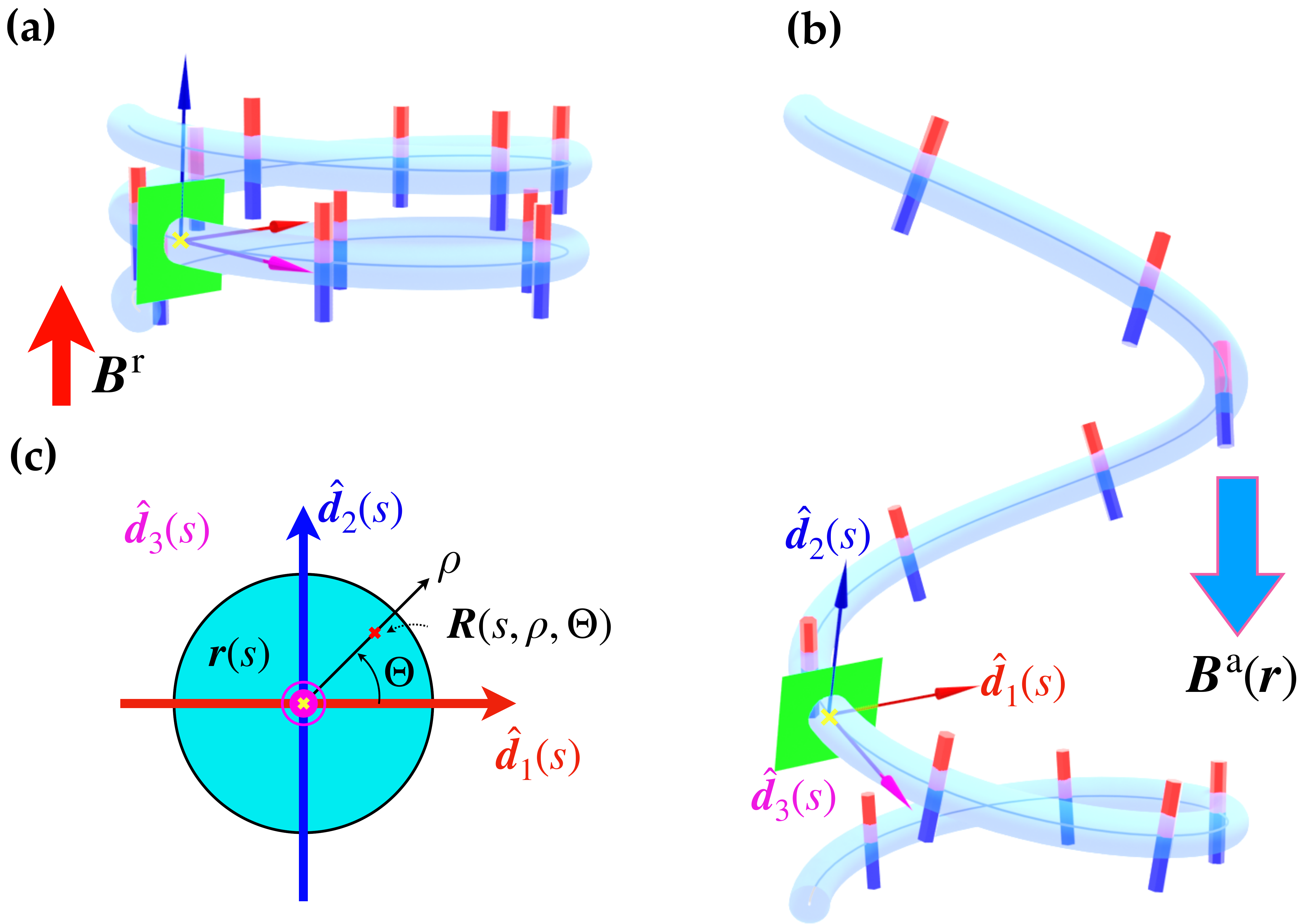}
    \caption{Definition of the problem. (a) A naturally curved and twisted rod is magnetized permanently along ${\bm B}^{\rm r}$. (b) Subsequently, when a magnetic field ${\bm B}^{\rm a}$ is applied, the rod deforms due to the induced magnetic force and torque. The centerline of the rod is described uniquely by the rotation of the Cosserat frame $(\hat{\bm{d}}_1,\hat{\bm{d}}_2,\hat{\bm{d}}_3)$ along the arc-length $s$. (c) The cross section of the rod is assumed circular. 
    Any material point of the 3D rod is identified by the cylindrical coordinates $(s,\rho,\Theta)$, where polar coordinates $(\rho,\Theta)$ are set on the $\hat{\bm{d}}_1-\hat{\bm{d}}_2$ plane for each value of $s$.}
    \label{fig:Schematic}
\end{figure}

In this section, we define the problem and identify the relevant variables. 
We consider a naturally curved and twisted (inextensible) rod of circular cross-section with diameter $d$ (area $A = \pi d^2/4$) and total length $L$ (Fig.~\ref{fig:Schematic}(a))~\cite{Landau:1980,Powers:2010cv,Audoly:2010Book,Bigoni:2015Book,Goriely:2017Book}. 
The goal of our study was to derive a reduced-order theoretical framework to describe the deformation of the centerline, ${\bm r}(s)$, of this hard magnetic rod, upon magnetic actuation. The arc-length of the centerline is $s$~$(0\leq s\leq L)$. 
In particular, we seek to derive the set of governing ordinary differential equations (ODEs); i.e., the strong form of the problem depicting force and moment balance of the system.
This section details the kinematics and constitutive relations of our system, in Sec.~\ref{sec:kin} and \ref{sec:mech}, respectively. 

\subsection{Kinematics}\label{sec:kin}

Our rod is slender such that the diameter is significantly smaller than its length; $d/L\ll1$.
Throughout this paper, we denote the quantities in the reference configuration with $\accentset{\circ}{({\cdot})}$. For example, we write the centerline position in the reference configuration as $\mr{\bm r}(s)$. The 2D orthogonal basis $\hat{\bm{d}}_1$-$\hat{\bm{d}}_2$ is defined on the cross-section normal to the tangent vector ${\hat{\bm{d}}}_3 = {\bm r}'(s)$, where $(\cdot)'$ represents differentiation with respect to $s$. 
With the help of the Cosserat frame basis ${\hat{\bm{d}}}_a~(a=1,2,3)$ defined here, one can uniquely identify the configuration of the rod centerline (other than rigid body translations and rotations). 
The rotation of the Cosserat frame can be represented using the Darboux vector ${\bm \Omega} = \Omega_a\hat{\bm{d}}_a$ as
\begin{eqnarray}
{\hat{\bm{d}}}_a ' &=& {\bm \Omega}\times{\hat{\bm{d}}}_a,
\label{eq:kinematic}
\end{eqnarray}
where $\Omega_a$ are the rotation rates of the Cosserat frame around $\hat{\bm{d}}_a$~\cite{Landau:1980,Powers:2010cv,Audoly:2010Book,Bigoni:2015Book,Goriely:2017Book}. We can then define the slenderness parameters as $\mr{\Omega}_ad$.

The 3D reference configuration of the rod can be described uniquely by introducing the cylindrical coordinates $\xi_{\mu} = s, \rho ,\Theta$ ($0\leq\rho\leq d/2$, $0\leq\Theta\leq2\pi$), whose origin corresponds to ${\bm r}(s)$ as shown in Fig.~\ref{fig:Schematic} (c). For example, the position vector of the 3D rod is 
\begin{eqnarray}
{\bm R}(s,\rho,\Theta) = {\bm r}(s) + \rho\cos\Theta~{\hat{\bm{d}}}_1(s) + \rho\sin\Theta~{\hat{\bm{d}}}_2(s). 
\label{eq:R}
\end{eqnarray}
Once the geometry is set up, we can obtain information about the deformation of the 3D body, which is characterized by the deformation gradient ${\bf F}$, defined as
\begin{eqnarray}
{\bf F} \equiv \frac{\partial {\bm R}}{\partial \mr{\bm R}} = {\bm t}_{\mu}\otimes\mr{\bm t}_{\mu}^{\dagger}.
\label{def:F}
\end{eqnarray}
The tangent vector along the curvilinear coordinate, ${\bm t}_{\mu}$, is expressed as ${\bm t}_{\mu} = \partial{\bm R}/\partial\xi_{\mu}$, where $\xi_{\mu} = s,\rho,\Theta$ for $\mu=1,2,3$, respectively. The reciprocal basis of ${\bm t}_{\mu}$ denoted by, ${\bm t}_{\mu} ^{\dagger}$, is defined as ${\bm t}_{\mu} ^{\dagger} \cdot {\bm t}_{\nu} = \delta_{\mu\nu}$ with the Kronecker delta $\delta_{\mu\nu}$, which is equivalently written as ${\bm t}_{\mu}^{\dagger} \equiv \partial\xi _{\mu} / \partial {\bm R}$. By using the kinematic relations, we can calculate the tangent vectors. 
Note, for example, that when $\mr{\bm t}_{\mu}^{\dagger}$ and ${\bm t}_{\mu}$ are expressed in terms of the Cartesian basis $\hat{\bm{e}}_{i}(i = x,y,z)$, each component of $\mathbf{F}$ is calculated as $F_{ij} = ({\bm t}_{\mu}\cdot\hat{\bm{e}}_{i})(\mr{\bm t}_{\mu}^{\dagger}\cdot\hat{\bm{e}}_{j})$. 
Throughout this paper, $a,b,c$ and $i,j,k$ are indices of the Cosserat frame and the Cartesian basis, respectively, if not specified. 

\subsection{Constitutive relation and Elasto-magnetic potential}\label{sec:mech}

We assume that the rod is made of a Hookean material, where the internal stress is linearly proportional to the strain~\cite{Audoly:2010Book}.  
The rod is permanently magnetized in the reference configuration with a magnetic flux density ${\bm B}^{\rm r}$. We investigate the deformation of the rod upon the application of an external applied magnetic flux density vector ${\bm B}^{\rm a}({\bm r})$ (Fig.~\ref{fig:Schematic}(b)). Note that the applied field $\bm{B}^{\rm a}$ is set to vary in space. We will specialize our theory in the cases of either homogeneous $\partial\bm{B}^{\rm a}/\partial r_i=0$ or inhomogenous fields $\partial\bm{B}^{\rm a}/\partial r_i\ne 0$, making use of the precision experimental framework introduced in section \ref{sec:apparatus}. 

As a starting point for the derivation of our Kirchhoff-like theory for hard magnetic rods, we will make use of the Helmholtz free energy for ideal hard-magnetic soft materials that was introduced recently in Refs.~\cite{Kim:2018bu,Zhao:2019hk}; which we review next. 
Once the rod is magnetized, all magnetized moments are parallel with $\bm{B}^{\rm r}$ (fully saturated). Hence, it is reasonable to assume that the magnetic permeability of the hard magnetic material is close to that of vacuum $\mu_0$. The magnetized moments will not affect the (surrounding) applied field $\bm{B}^{\rm a}$ upon actuation. 
Besides, the permeability of the elastomer that we will use in the experiments (section \ref{sec:apparatus}) is nearly the same as that of the vacuum $\mu_0$.
The magnetization density vector per unit volume in the deformed configuration is modeled as the ``rotation" of ${\bm B}^{\rm r}$ as $\bm{\mathcal M}_{\rm 3D} = {\bf F}{\bm B}^{\rm r}/\mu_0$. Thus, the total elasto-magnetic energy for a hard magnetic rod can be simplified as the work to align the residual magnetic moment of the material with ${\bm B}^a$:
\begin{eqnarray}
{E}_{\rm mag} = -\iiint{\bm {\mc M}}_{\rm 3D} {\bm B}^{\rm a} \rho d\rho d\Theta ds = -\iiint\frac{1}{\mu_0} ({\bf F}{\bm B}^{\rm r})\cdot{\bm B}^{\rm a}\rho d\rho d\Theta ds\,.
\label{eq:Wmag3D}
\end{eqnarray}
Note here that we do not consider the internal stress or long-range forces induced by a dipole-dipole interaction~\cite{Hall:2013gm,Jackson:100964,Vella:2014ix,Gonzalez2021}. The validity of this simplifying assumption will be evaluated thoroughly in Sec.~\ref{sec:Limitation}.
In the following, by integrating Eq.~(\ref{eq:Wmag3D}) over the cross-section, we seek to obtain the reduced expression for the elasto-magnetic energy density of the rod per a unit {\it length}, up to the leading order in the slenderness. 

\section{Theory of a Kirchhoff-like equation for hard magnetic elastic rods}
\label{sec:Theory}

In this section, we derive the equilibrium equations (ODEs) describing the geometrically nonlinear deformation of hard magnetic rods. The main results of this section are Eqs.~(\ref{eq:Feq}) and (\ref{eq:Meq}), that is the magnetic Kirchhoff rod equations. To arrive at this set of ODEs, in Sec.~\ref{sec:reduction}, we will follow the procedure of dimensional reduction. 
In Sec.~\ref{sec:review}, we review the derivation of the {\it non-magnetic} Kirchhoff rod equations based on the principle of virtual work. 
We derive the elasto-magnetic force $\bm{p}_{\rm mag}$ and torque $\bm{q}_{\rm mag}$ through variational calculus in Sec.~\ref{sec:vw}, thus obtaining the magnetic Kirchhoff rod equations for hard magnetic rods.
In Sec.~\ref{sec:Reduc2D}, we show that our framework reproduces previous works on the planar deformation of hard magnetic beams and elastica~\cite{Lum:2016fc,Ciambella:2020je,Wang:2020du,Yan:2020prep}, whereas in Sec.~\ref{sec:3D} we will employ our framework to study cases where the deformation is three-dimensional.

\subsection{Reduced elasto-magnetic and elastic energy}\label{sec:reduction}
Integrating Eq.~(\ref{eq:Wmag3D}) over the cross-section will yield the 1D reduced energy of the rod system, to leading order of slenderness $O(1)$ ($\mr{\Omega}_ad,\Omega_ad\ll 1$). 
To proceed, we first expand the deformation gradient as $\mathbf{F}(s,\rho,\Theta) = \mathbf{D}(s) + O(\mr{\Omega}_ad,\Omega_ad)$, where
\begin{eqnarray}
\mathbf{D}(s) \equiv {\hat{\bm{d}}}_a(s)\otimes\mr{\hat{\bm{d}}}_a(s)\label{def:D}
\end{eqnarray}
is the reduced deformation gradient, mapping the undeformed tangent space to the deformed tangent space along the centerline of the rod.
The deformation gradient is now reduced to the tensor product between deformed and undeformed directors, as given by Eqs.~(\ref{def:F}) and (\ref{def:D}). 
Given that $\mathbf{D}$ is not a function of $\rho$ nor $\Theta$, the integral along the cross-section is replaced by the constant cross-section area $A$. Hence, from Eq.~\eqref{eq:Wmag3D}, we obtain the reduced elasto-magnetic energy functional as
\begin{eqnarray}
{E}_{\rm mag} = -\int_0 ^L\bm{\mathcal{M}}(s)\cdot{\bm B}^{\rm a}({\bm r}(s))ds\,,
\label{eq:WmagTot}
\end{eqnarray}
where we have introduced the magnetization density vector per a unit length $\bm{\mathcal{M}}$ (by contrast to the 3D counterpart, $\bm {{\mc M}}_{\rm 3D}$, introduced above) defined as
\begin{eqnarray}
\bm{\mathcal{M}}(s) \equiv \frac{{A}}{\mu_0} ({\bf D}(s){\bm B}^{\rm r}).
\label{eq:MagDef}
\end{eqnarray}
It is important to note that, when the applied field ${\bm B}^{\rm a}$ is not uniform, ${\bm B}^{\rm a}$ at the material point $\bm{R}(s,\rho,\Theta)$ is represented by the value at the centerline position ${\bm r}(s)$ as $\bm{B}^{\rm a}(\bm{R}(s,\rho,\Theta))\simeq\bm{B}^{\rm a}(\bm{r}(s))$. 
The spatial variation of ${\bm B}^{\rm a}$ along the cross section is at higher order of slenderness $O(\mr{\Omega}_ad, \Omega_ad)$ as $\bm{B}^{\rm a}(\bm{R})=\bm{B}^{\rm a}(\bm{r}) + O(\mr{\Omega}_ad, \Omega_ad)$, and can, therefore, be neglected.

We now want to derive the equilibrium equations, augmenting the non-magnetic Kirchhoff rod equations. In the absence of magnetic effects, we recall that, from the classic literature \cite{Audoly:2010Book,Landau:1980,Powers:2010cv,Bigoni:2015Book,Goriely:2017Book}, the total elastic energy for a naturally curved and twisted elastic (Hookean) rod in the absence of a magnetic field is given by
\begin{eqnarray}
{E}_{\rm el} = \int_0 ^L\frac{EI}{2} \Bigl\{\Bigl(\Omega_1 - \mr{ \Omega}_1\Bigr)^2 + \Bigl(\Omega_2 - \mr{ \Omega}_2\Bigr)^2\Bigr\} + \frac{G J}{2}\Bigl(\Omega_3 - \mr{ \Omega}_3\Bigr) ^2 ds,
\end{eqnarray}
where $EI$ and $G J$ are the bending and twist moduli, respectively, with $E$ as the Young modulus, $G = E/\{2(1+\nu)\}$ as the shear modulus with the Poisson ratio $\nu$, $I = \pi d^4/64$ as the second moment of area and $J = \pi d^4/32$ as the torsional constant.

The set of equations of mechanical equilibrium is derived through the principle of virtual work as
\begin{eqnarray}
\delta E_{\rm el} + \delta {E}_{\rm mag} = \delta W_{\rm ex},
\label{eq:PVW}
\end{eqnarray}
where $W_{\rm ex}$ is the work done by the external forces and torques, $({\bm F}_{\rm ex}(0), {\bm M}_{\rm ex}(0))$ and $({\bm F}_{\rm ex}(L), {\bm M}_{\rm ex}(L))$, applied at $s = 0$ and $s=L$, respectively, together with the external forces $\bm{p}(s)$ and torques $\bm{q}(s)$ per unit length (e.g., gravity, viscous force and torque). 

The magnetic Kirchhoff rod equations can be derived directly from Eq.~(\ref{eq:PVW}), which will be tackled in Sec. \ref{sec:vw}. 
Before describing our theory for magnetic rods, we review the well-established and classic ({\it non-magnetic}) Kirchhoff rod equations.
Note that $\delta E_{\rm el} = \delta W_{\rm ex}$ corresponds to the classic case in the absence of the magnetic field.
Given that $E_{\rm mag}$ is additive in the total energy, the variation of $E_{\rm mag}$ alone will yield the expressions of the elasto-magnetic force $\bm{p}_{\rm mag}$ and torque $\bm{q}_{\rm mag}$ per unit length, which must be added to the force and moment balance equations as an external force and torque, respectively.


\subsection{Review of the (non-magnetic) Kirchhoff rod equations}\label{sec:review}

In the previous section, we have derived the reduced expressions for the elasto-magnetic potential. In the absence of the magnetic field $E_{\rm mag} = 0$, the variations of $E_{\rm el}$ and $W_{\rm ex}$ follow the classic (non-magnetic) Kirchhoff rod equations. In this section, for completeness, we review their classic derivation.
The Kirchhoff rod equations can be derived by considering the restricted variation of the total energy functional such that the centerline does not stretch: $\delta(ds) = 0$ (i.e., the differentiation with respect to $s$, $d/ds$, and the variation, $\delta$, commute). A detailed account of this derivation can be found in Ref.~\cite{Powers:2010cv}; in what follows, we provide just a brief sketch.

The variation of the elastic energy functional $\delta E_{\rm el}$ is written using the variation of $\Omega_a$, $\delta\Omega_a$:
\begin{eqnarray}
\delta E_{\rm el} &=& \int_0 ^L EI(\Omega_1-\mr{\Omega}_1)\delta\Omega_1 + EI(\Omega_2-\mr{\Omega}_2)\delta\Omega_2 + GJ(\Omega_3-\mr{\Omega}_3)\delta\Omega_3ds.
\label{eq:dEel0}
\end{eqnarray}
To rewrite Eq.~(\ref{eq:dEel0}) with respect to the angular parameters and infinitesimal angles $\bm{\epsilon}$~\cite{Powers:2010cv}, the infinitesimal rotation vector of the Cosserat frame ${\bm \epsilon}$ is introduced as ${\bm \epsilon} \equiv \delta\chi_a\hat{\bm{d}}_a$ whose components are expressed as $\delta\chi_1 \equiv (\delta\hat{\bm{d}}_2)\cdot\hat{\bm{d}}_3 = -(\delta {\bm r})' \cdot\hat{\bm{d}}_2$, $\delta\chi_2 \equiv (\delta\hat{\bm{d}}_3)\cdot\hat{\bm{d}}_1 =  (\delta {\bm r})' \cdot\hat{\bm{d}}_1$, and $\delta\chi_3 \equiv (\delta\hat{\bm{d}}_1)\cdot\hat{\bm{d}}_2 = -(\delta\hat{\bm{d}}_2)\cdot\hat{\bm{d}}_1$. 
The end-forces and moments are applied at $s = 0$ and $s = L$ as ${\bm F}_{\rm ex}(0), {\bm M}_{\rm ex}(0), {\bm F}_{\rm ex}(L)$, and ${\bm M}_{\rm ex}(L)$, respectively. The rod is subjected to the external force and torque per unit length ${\bm p}(s)$ and ${\bm q}(s)$. 
The variation of the work done by the external forces and torques (in the absence of magnetic fields) is
\begin{eqnarray}
\delta W_{\rm ex} &=& {\bm F}_{\rm ex}(0)\cdot\delta{\bm r}(0) + {\bm F}_{\rm ex}(L)\cdot\delta{\bm r}(L) + {\bm M}_{\rm ex}(0)\cdot{\bm \epsilon}(0) + {\bm M}_{\rm ex}(L)\cdot{\bm \epsilon}(L)\nonumber\\
&&+ \int_0 ^L \left\{{\bm p}(s)\cdot\delta{\bm r}(s) + {\bm q}(s)\cdot{\bm \epsilon}(s)\right\}ds.
\label{eq:dWex}
\end{eqnarray}
The internal and external virtual works expressed with respect to $\delta{\bm r}$ and ${\bm \epsilon}$ are then equated as $\delta E_{\rm el} = \delta W_{\rm ex}$.
Collecting the terms associated with $\delta\bm{r}(s)$ and $\bm{\epsilon}(s)$ yields the (non-magnetic) Kirchhoff rod equations
\begin{eqnarray}
&&{\bm F}'(s) + {\bm p}(s) = 0\label{eq:Feq0},\\
&&{\bm M}' (s)+ \hat{\bm{d}}_3(s)\times{\bm F}(s) +{\bm q}(s) = 0\label{eq:Meq0},
\end{eqnarray}
with the internal moment vector ${\bm M}(s) = M_a\hat{\bm{d}}_a = EI\{(\Omega_1 - \mr{\Omega}_1)\hat{\bm{d}}_1 + (\Omega_2 -  \mr{\Omega}_2)\hat{\bm{d}}_2\} + G J(\Omega_3 - \mr{\Omega}_3)\hat{\bm{d}}_3$, and $\bm{F}(s)$ is the internal force acting over the cross section at $s$. 
Equations (\ref{eq:Feq0}) and (\ref{eq:Meq0}) are the equilibrium equations of forces and moments for the rod, respectively.
The boundary conditions can be also derived from variational calculus~\cite{Powers:2010cv} as
\begin{eqnarray}
{\bm M}(0) = -{\bm M}_{\rm ex}(0),~~{\bm M}(L) = {\bm M}_{\rm ex}(L),~~
{\bm F}(0) = -{\bm F}_{\rm ex}(0),~~{\bm F}(L) = {\bm F}_{\rm ex}(L).
\label{eq:bc}
\end{eqnarray}

Next, we will tackle the derivation of the magnetic Kirchhoff rod equations based on the same procedure sketched above, while the elastic terms Eqs.~(\ref{eq:Feq0}) and (\ref{eq:Meq0}) remain unchanged. The last task is to compute the variation of the elasto-magnetic potential, $E_{\rm mag}$, with respect to $\delta\bm{r}(s)$ and $\bm{\epsilon}(s)$, $\delta E_{\rm mag}$, which is addressed next.

\subsection{Derivation of the magnetic Kirchhoff rod equations using the principle of virtual work}
\label{sec:vw}

We now set out to derive the magnetic Kirchhoff rod equations. First, we will compute the variation of the elasto-magnetic potential, $\delta E_{\rm mag}$. 
Equating the total internal and external virtual works as Eq.~(\ref{eq:PVW}), we will obtain the magnetic Kirchhoff rod equations. Given that $\delta E_{\rm el}, \delta W_{\rm ex}$, and $\delta E_{\rm mag}$ are additive in Eq.~(\ref{eq:PVW}), the variation of the elasto-magnetic potential $\delta E_{\rm mag}$ provides the elasto-magnetic force $\bm{p}_{\rm mag}$ and torque $\bm{q}_{\rm mag}$ explicitly, which will be readily added to Eqs.~(\ref{eq:Feq0}) and (\ref{eq:Meq0}) to capture magnetic effects.

Before calculating $\delta E_{\rm mag}$, we compute the variation of the Cosserat frame basis $\hat{\bm{d}}_a$, $\delta\hat{\bm{d}}_a$. 
Given that ${\bm \epsilon}$ represents the infinitesimal rotation around $\hat{\bm{d}}_a$, the variation of $\hat{\bm{d}}_a$ is $\delta\hat{\bm{d}}_a = {\bm \epsilon}\times\hat{\bm{d}}_a=-\varepsilon_{abc} \delta\chi_b\hat{\bm{d}}_c$, where $\varepsilon_{abc}$ is the Eddington epsilon. 
We aim to express $\delta E_{\rm mag}$ with respect to $\delta{\bm r}$ and ${\bm \epsilon}$. 
Note that $\delta\bm{B}^{\rm r}$ and $\delta\mr{\hat{\bm{d}}}_a$ are zero, since they do not depend on the state variables. 
We first determine the variation of the reduced deformation gradient ${\bf D}$ for both $\delta{\bm r}$ and ${\bm \epsilon}$ as
\begin{eqnarray}
\delta{{\bf D}} &=& \delta(\hat{\bm{d}}_a\otimes\mr{\hat{\bm{d}}}_a)
= (\delta\hat{\bm{d}}_a)\otimes\mr{\hat{\bm{d}}}_a
\nonumber\\
&=& \{(\delta{\bm r} ' \cdot\hat{\bm{d}}_2) \}\mathbf{\Delta}_1
-\{(\delta{\bm r} ' \cdot\hat{\bm{d}}_1) \}\mathbf{\Delta}_2-\{({\bm \epsilon} \cdot\hat{\bm{d}}_3) \}\mathbf{\Delta}_3\nonumber\\
&=&- \{({\bm \epsilon} \cdot\hat{\bm{d}}_a) \}\mathbf{\Delta}_a\,,
\label{eq:dD2}
\end{eqnarray}
where we have defined the new tensor $\mathbf{\Delta}_a~(a=1,2,3)$ associated with the magnetic torque as
\begin{eqnarray}
\mathbf{\Delta}_a(s) \equiv \varepsilon_{abc}\hat{\bm{d}}_b(s) \otimes\mr{\hat{\bm{d}}}_c(s).
\end{eqnarray}
To obtain Eq.~(\ref{eq:dD2}), we have used the orthogonality of the basis $\hat{\bm{d}}_b\times\hat{\bm{d}}_c = \varepsilon_{abc}\hat{\bm{d}}_a$. We also used the fact that, for an arbitrary vector ${\bm a} = {\bm a}(s)$, $\delta{\bm r}'\cdot{\bm a} = \delta(\hat{\bm{d}}_3)\cdot{\bm a}=({\bm \epsilon}\times\hat{\bm{d}}_3)\cdot{\bm a} = {\bm \epsilon}\cdot(\hat{\bm{d}}_3\times{\bm a})$ holds, where we set ${\bm a} = \hat{\bm{d}}_1$ or ${\bm a} = \hat{\bm{d}}_2$. Given that the variation of ${\bm B}^{\rm a}(\bm{r})$ acts as $\delta {\bm B}^{\rm a} = (\partial {\bm B}^{\rm a}/\partial r_i)\delta r_{i} \equiv (\nabla {\bm B}^{\rm a})\delta{\bm r}$, we compute the variation of $E_{\rm mag}$, $\delta E_{\rm mag}$ as
\begin{eqnarray}
\delta E_{\rm mag} &=& - \int_0 ^L [\delta{\bm r}\cdot{\bm p}_{\rm mag}(s) + {\bm \epsilon}\cdot{\bm q}_{\rm mag}(s)] ds
\label{eq:dWmag},\\
{\bm p}_{\rm mag}(s) &\equiv& \frac{A}{\mu_0}({\bf D}(s){\bm B}^{\rm r})\cdot(\nabla {\bm B}^{\rm a}) = \bm{\mathcal{M}}\cdot(\nabla {\bm B}^{\rm a}),\label{eq:p_def}\\
{\bm q}_{\rm mag}(s) &\equiv& -\frac{A}{\mu_0}\left\{(\mathbf{\Delta}_a(s){\bm B}^{\rm r})\cdot{\bm B}^{\rm a}\right\}\hat{\bm{d}}_a(s) = \bm{\mathcal{M}}\times{\bm B}^{\rm a}.
\label{eq:q_def}
\end{eqnarray}
We have introduced ${\bm p}_{\rm mag}$ and ${\bm q}_{\rm mag}$ as the elasto-magnetic force and torque, respectively. 
When the applied field is homogeneous $\partial\bm{B}^{\rm a}/\partial r_i = 0$, the magnetic torque $\bm{q}_{\rm mag} \ne 0$ drives the deformation (i.e., $\bm{p}_{\rm mag} = 0$), while both $\bm{p}_{\rm mag}$ and $\bm{q}_{\rm mag}$ contribute to the deformation if the applied field is inhomogeneous. 

Adding the elasto-magnetic force and torque terms from Eqs.~(\ref{eq:p_def}) and (\ref{eq:q_def}) into the classic (non-magnetic) Kirchhoff rod equations (\ref{eq:Feq0}) and (\ref{eq:Meq0}), yields the force and moment balance equations:
\begin{eqnarray}
&&{\bm F}'(s) + {\bm p}(s) + {\bm p}_{\rm mag}(s) = 0\label{eq:Feq},\\
&&{\bm M}' (s)+ \hat{\bm{d}}_3(s)\times{\bm F}(s) +{\bm q}(s)+ {\bm q}_{\rm mag}(s) = 0\label{eq:Meq},
\end{eqnarray}
where the magnetic force ${\bm p}_{\rm mag}$ and torque ${\bm q}_{\rm mag}$ were defined in Eqs.~(\ref{eq:p_def}) and (\ref{eq:q_def}), respectively. 
Note that the boundary conditions specified in Eq.~(\ref{eq:bc}) remain unchanged in the elasto-magnetic case because they do not appear in Eq.~(\ref{eq:dWmag}).
We refer to the set of Eqs. (\ref{eq:Feq}) and (\ref{eq:Meq}) as the {\it magnetic Kirchhoff rod equations}, where the coupling between elasticity and magnetism is captured by ${\bm p}_{\rm mag}$ and ${\bm q}_{\rm mag}$.

\subsection{Reduction of the magnetic Kirchhoff rod equations to planar deformations}\label{sec:Reduc2D}

In this subsection, for verification purposes, we show that the magnetic Kirchhoff equations derived above can reproduce existing results in the literature for the hard magnetic beams and elastica~\cite{Wang:2020du, Yan:2020prep}, loaded under a constant or constant gradient magnetic fields. 
In Ref.~\cite{Wang:2020du}, a theoretical model under constant field was derived within the continuum framework and validated against experiments. 
The governing equations for a more general description that can tackle either under constant or constant gradient fields were developed through dimensional reduction and validated experimentally in Ref.~\cite{Yan:2020prep}.

Consider a naturally straight rod clamped at $s = 0$ (${\bm r}(0)={\bm 0}$), while the other end at $s = L$ is set to be force and momentum free. The rod is free from any external forces and torques other than magnetic fields, that is ${\bm p} = {\bm q} = 0$. We will derive the governing equation for the bending angle $\theta(s)$ in the $x$-$y$ plane such that $\theta(s) = 0$ holds in the reference configuration. The constitutive law is ${\bm M}(s) = EI\theta'(s)\hat{\bm{e}}_z$.
We will study two cases: a naturally straight hard magnetic rod loaded under a constant or constant-gradient fields, which have been studied in Ref.~\cite{Wang:2020du} and Ref.~\cite{Yan:2020prep}, respectively.

\begin{figure}[!ht]
    \centering
    \includegraphics[width=1.0\textwidth]{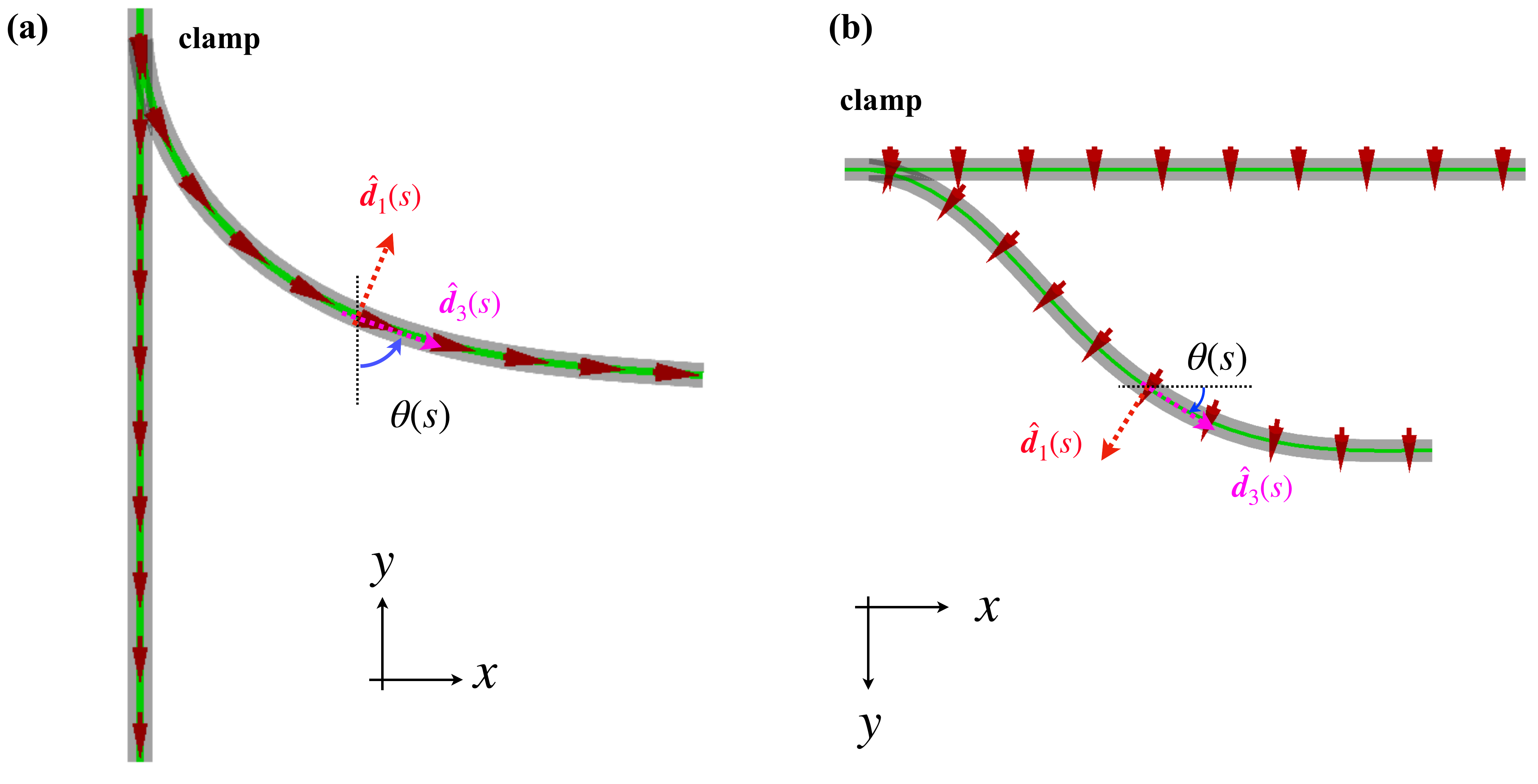}
    \caption{Reduction of magnetic Kirchhoff rod equations to planar (2D) cases. 
    (a) Schematics of a hard magnetic rod under constant field that deforms in 2D. The rod is magnetized along the tangent of the rod $\hat{\bm{d}}_3$.
    (b) Schematics of a hard magnetic rod under constant {\it gradient} field. The rod is magnetized along $\hat{\bm{d}}_1$.}
    \label{fig:planar}
\end{figure}

First, we consider the case of a rod under a {\it constant} magnetic field ${\bm B}^{\rm a} = B^{\rm a}\hat{\bm{e}}_y$; see~Fig.~\ref{fig:planar}(a). The rod is magnetized as ${\bm B}^r = -B^r\hat{\bm{e}}_y$ along the center-line ( $\mr{\hat{\bm{d}}}_1 = \hat{\bm{e}}_x$, $\mr{\hat{\bm{d}}}_2= \hat{\bm{e}}_z$, $\mr{\hat{\bm{d}}}_3 = -\hat{\bm{e}}_y$) with $B^r > 0$ and $B^{\rm a} > 0$. 
In the absence of the applied field, the rod is naturally straight along $\hat{\bm{e}}_y$ as $\mr{\hat{d}}_3 = - \hat{\bm{e}}_y$. The Cosserat frame in the reference configuration is given by $\hat{\bm{d}}_1 = (\cos\theta, \sin\theta, 0)$, $\hat{\bm{d}}_2 = \hat{\bm{e}}_z$ and $\hat{\bm{d}}_3 = (\sin\theta, -\cos\theta, 0)$. 
Since the applied field is homogeneous, we have ${\bm p}_{\rm mag} = 0$ and $\bm{F}(s) = 0$. 
To calculate ${\bm q}_{\rm mag}$ in Eq.~(\ref{eq:q_def}), we first need to express $\bm{\mathcal{M}}$ with respect to $\theta$. From Eq.~(\ref{eq:MagDef}), $\bm{\mathcal{M}}$ is simplified as
\begin{eqnarray}
\bm{\mathcal{M}} = \frac{AB^{\rm r}}{\mu_0}\hat{\bm{d}}_3.
\label{eq:MagPlanar1}
\end{eqnarray}
Substituting ${\bm F}(s) = 0$ and Eq.~(\ref{eq:MagPlanar1}) into the moment balance equation (Eq.~(\ref{eq:Meq})), we can reproduce the equation for the hard magnetic elastica~\cite{Wang:2020du} as
\begin{eqnarray}
EI\theta''(s) + \frac{AB^{\rm r} B^{\rm a}}{\mu_0}\sin\theta(s) = 0,
\label{eq:HardMagElastica}
\end{eqnarray}
with the boundary conditions $\theta(0) = 0$ and $\theta'(L) = 0$. 
As discussed in Ref.~\cite{Wang:2020du}, Eq.~(\ref{eq:HardMagElastica}) is mathematically equivalent to the governing equation of the clamped-free elastica. In this configuration, the magnetic rod is known to buckle at the critical applied field~\cite{Wang:2020du}
\begin{eqnarray}
B^{{\rm a}*} _{\rm bend} =\left(\frac{\pi}{2}\right)^2\frac{\mu_0 EI}{AB^rL^2}\,,
\label{eq:Bbend}
\end{eqnarray}
a result that will be used in Sec.~\ref{sec:HelixBuckling} to connect our 3D analytical results with the existing 2D results in literature~\cite{Wang:2020du}.

Second, we consider the case of a rod in a {\it constant gradient} magnetic field: ${\bm B}^{\rm a}({\bm r}) =  by \hat{\bm{e}}_y$, where $b$ is a constant. 
The rod is clamped along $\hat{\bm{e}}_x$ ($y=0$) and magnetized as ${\bm B}^{\rm r} = B^{\rm r}\hat{\bm{e}}_y$~(Fig.~\ref{fig:planar} (b)). The Cosserat frame in the reference configuration is given by $\mr{\hat{\bm{d}}}_1 = \hat{\bm{e}}_y$, $\mr{\hat{\bm{d}}}_2 = \hat{\bm{e}}_z$, and $\mr{\hat{\bm{d}}}_3 = \hat{\bm{e}}_x$. We can write the Cosserat frame in the deformed configuration as ${\hat{\bm{d}}}_1 = (-\sin\theta, \cos\theta, 0)$, $\hat{\bm{d}}_2 = \hat{\bm{e}}_z$, and ${\hat{\bm{d}}}_3 = (\cos\theta, \sin\theta, 0)$. 
Hence, the magnetization vector is simplified from Eq.~(\ref{eq:MagDef}) as
\begin{eqnarray}
\bm{\mathcal{M}} = \frac{AB^{\rm r}}{\mu_0}\hat{\bm{d}}_1.
\label{eq:MagPlanar2}
\end{eqnarray}
Given that the rod is in a constant gradient field, $\bm{p}_{\rm mag}$ is non-zero. From Eqs.~(\ref{eq:p_def}) and (\ref{eq:MagPlanar2}), we find $\bm{p}_{\rm mag}(s) = (AB^{\rm r}b/\mu_0)\cos\theta(s)~\hat{\bm{e}}_y$. 
The force balance in Eq.~(\ref{eq:Feq}) can now be integrated as
\begin{eqnarray}
{\bm F}(s) &=& {\bm F}(L) + \int_s ^L {\bm p}_{\rm mag}(s')ds'\nonumber\\
&=& \frac{AB^rb}{\mu_0}\hat{\bm{e}}_y\int_s ^L \cos\theta(s')ds'.
\end{eqnarray}
Plugging this result into the moment balance in Eq. \eqref{eq:Meq}, we obtain the following integro-differential equation
\begin{eqnarray}
EI\frac{d^2{\theta}}{d{s}^2} - \frac{AB^r b}{\mu_0} \left\{{y}({s})\sin{\theta}({s}) -\cos{\theta}({s})\int_{{s}} ^1 \cos{\theta}(s') ds'\right\} = 0,\label{eq:HeavyMagElastica}
\end{eqnarray}
whose second and third terms originate from magnetic torque and force, respectively. 
Equation (\ref{eq:HeavyMagElastica}) has been recently derived and validated experimentally in Refs.~\cite{Wang:2020du,Yan:2020prep}, verifying our reduced framework presented in this section.

We note that the inhomogeneous applied field can be interpreted as a distributed load on the magnetic rod. 
Indeed, in the limit of small bending angle $\theta\ll1$, Eq. (\ref{eq:HeavyMagElastica}) is identical to that of an elastica under gravity~\cite{Wang:1986kl}. 
By analogy with the gravitational case, we can introduce the {\it magneto-bending} length $\ell_m$ as
\begin{eqnarray}
\ell_m \equiv \left(\frac{\mu_0 EI}{AB^{\rm r}b}\right)^{1/3},
\label{eq:magnetobending}
\end{eqnarray}
which can be regarded as a (persistence) length quantifying the relative importance of the magnetic and elastic effects. When $\lambda_m \equiv L/\ell_m\ll1$, the gradient of the magnetic field is negligible, while in the case of $\lambda_m\gg1$, the magnetic gradient plays an important role in the bending of the magnetic rod.

We have shown that the magnetic Kirchhoff rod equations derived in Sec.~\ref{sec:vw} are able to reproduce existing results reported in Refs.~\cite{Wang:2020du,Yan:2020prep} for the planar (2D) deformation of the magnetic rod. 
The specialization to the 2D cases verifies our theory, even if only partially. 
In the following sections, we will combine discrete simulation and precision experiments to study non-planar (3D) deformations of the rod, towards the validation of our theoretical framework.

\section{Discrete simulations of magneto-elastic rods}
\label{sec:Sim}
Although our magnetic Kirchhoff rod model reproduces the governing equation of planar hard magnetic beams and elastica, computing the 3D nonlinear geometric deformation requires, in general, a numerical approach. Specifically, we discretize the rod-centerline as a set of connected particles via straight rigid segments~\cite{Chirico:1994hm,Wada:2007dt,Wada:2007kw,Vogel:2010gc,Reichert:2006Thesis,Bergou:2008dt,Morigaki:2016dp,Sano:2017ki,Sano:2018ji,Sano:2019hk}. This discrete method is widely used to simulate the large deformation of rod-like structures such as nano-springs~\cite{Wada:2007dt}, DNA~\cite{Chirico:1994hm}, helical bacteria~\cite{Wada:2007kw}, flagella~\cite{Reichert:2006Thesis,Vogel:2010gc}, human hairs~\cite{Bergou:2008dt}, and gift-wrapping ribbons~\cite{Morigaki:2016dp,Sano:2017ki,Sano:2018ji,Sano:2019hk}.
Within this computational framework, the rod centerline is regarded as the chain of straight segments of natural length $\ell_0$ and an equal mass $m = \rho A L/N$. At each node $i = 1, 2,...,N$, we assigned a discrete Cosserat frame basis vectors $(\hat{\bm{d}}_{1,i}, \hat{\bm{d}}_{2,i}, \hat{\bm{d}}_{3,i})$, corresponding to $\hat{\bm{d}}_a$ in the continuum model. The elastic bending and twist deformations are represented by the Euler angles between adjacent discretized Cosserat frame basis vectors~\cite{Chirico:1994hm,Wada:2007dt,Wada:2007kw,Reichert:2006Thesis,Vogel:2010gc,Bergou:2008dt,Morigaki:2016dp,Sano:2017ki,Sano:2018ji,Sano:2019hk}.

In the present section, we detail the simulation method, with a particular focus on the discrete versions for the elasto-magnetic forces, $\bm{\mathcal{P}}$, and torques, $\mathcal{Q}$, in Sec.~\ref{sec:EM_sim}. The equations of motion for the discrete segments follow in Sec.~\ref{sec:EoM_sim}.

\subsection{Elasto-magnetic force and torque in the discrete simulation}\label{sec:EM_sim}

To compute the nonlinear geometric deformation of our rod-like magnetic structures, we adopt dynamic simulations, instead of solving the static equilibrium equations. 
Static analysis packages for ODEs (e.g., AUTO~\cite{Doedel:AUTO}) might have been able to compute the deformation of our MRE rods. However, it is not trivial to implement the elasto-magnetic interactions into such packages due to the geometric non-linearity of our MRE rods. 
To facilitate numerical convergence to the mechanical equilibrium, we perform the dynamic simulation by including the inertial terms in Eqs.~(\ref{eq:Feq}) and (\ref{eq:Meq}). 
Specifically, we will derive the equations of motion for ${\bm r}$ and the twist angle $\chi_3$ (see Sec.~\ref{sec:review} for the definition of $\chi_3$). We recall that the expressions of the elasto-magnetic force ${\bm p}_{\rm mag}$ and torque ${\bm q}_{\rm mag}$, provided in Eqs.~(\ref{eq:dWmag})-(\ref{eq:q_def}), have been derived from the continuum framework, through the variation $\delta E_{\rm mag}$ with respect to the set of $(\delta{\bm r}, {\bm \epsilon})$. 
The set of infinitesimal bending and twist angles were represented by ${\bm \epsilon} = \delta\chi_a\hat{\bm{d}}_a$ as in Sec.~\ref{sec:review}.
In the discrete simulation framework~\cite{Chirico:1994hm,Wada:2007dt,Wada:2007kw,Vogel:2010gc,Reichert:2006Thesis,Bergou:2008dt,Morigaki:2016dp,Sano:2017ki,Sano:2018ji,Sano:2019hk}, it is sufficient to derive the equation of motion of the centerline position and the twist angle $(\bm{r},\chi_{3})$ alone.
In other words, the remaining bending angles $\chi_{1}$ and $\chi_{2}$ are determined from the centerline positions $\bm{r}$, as we explain below~\cite{Powers:2010cv}.
To include the elasto-magnetic force and torque in the equations of motion for $(\bm{r},\chi_{3})$, we will introduce the discrete version of $({\bm p}_{\rm mag}, {\bm q}_{\rm mag})$, as $(\bm{\mc{P}}, {\mc{Q}})$, by rewriting $\delta E_{\rm mag}$ (Eq.~(\ref{eq:dWmag})) with respect to the set of $(\bm{r},\chi_{3})$. 
The discrete version of the elasto-magnetic torque ${\mc{Q}}$ (a scalar) will correspond to the elasto-magnetic torque around $\hat{\bm{d}}_3$, while the remaining components of $\bm{q}_{\rm mag}$ (\textit{i.e.}, ${q}_{\rm mag1}$ and ${q}_{\rm mag2}$) will be included in the equation of motion for $\bm{r}$, thereby defining $\bm{\mc{P}}$, as we detail below.

Expressing $\delta\chi_1$ and $\delta\chi_2$ from Eq.~(\ref{eq:dWmag}) with the aid of $\delta\chi_1 = - (\delta {\bm r})'\cdot\hat{\bm{d}}_2$ and $\delta\chi_2 = (\delta {\bm r})'\cdot\hat{\bm{d}}_1$, we find the following equation after the partial integration of $(\delta {\bm r})'$:
\begin{eqnarray}
\delta E_{\rm mag}&=&- \int_0 ^L [\delta{\bm r}(s)\cdot{\bm p}_{\rm mag}(s) + {\bm \epsilon}(s)\cdot{\bm q}_{\rm mag}(s)] ds\nonumber\\
&=&- \int_0 ^L [\delta{\bm r}\cdot{\bm p}_{\rm mag} + \delta\chi_1q_{\rm mag1} + \delta\chi_2q_{\rm mag2} + \delta\chi_3q_{\rm mag3}] ds\nonumber\\
&=&- \int_0 ^L [\delta{\bm r}\cdot{\bm p}_{\rm mag} + \{- (\delta {\bm r})'\cdot\hat{\bm{d}}_2\}q_{\rm mag1} + \{(\delta {\bm r})'\cdot\hat{\bm{d}}_1\}q_{\rm mag2} + \delta\chi_3q_{\rm mag3}] ds\nonumber\\
&=& - \int_0 ^L \left[\delta{\bm r}\cdot\bm{\mc{P}}+ \delta\chi_3\mc{Q}\right]ds+ \left[\delta{\bm r}\cdot\left\{{q}_{{\rm mag}1}\hat{\bm{d}}_2-{q}_{{\rm mag}2}\hat{\bm{d}}_1\right\}\right]_0 ^L,
\end{eqnarray}
where we introduce the $a (=1,2,3)$-th component of the magnetic torque $q_{{\rm mag}a} = {\bm q}_{\rm mag}\cdot\hat{\bm{d}}_a$.
Here, we define the discrete version of the elasto magnetic force and torque as
\begin{eqnarray}
&&\bm{\mc{P}}\equiv {\bm p}_{\rm mag} + \left({q}_{{\rm mag}1}\hat{\bm{d}}_2 - {q}_{{\rm mag}2}\hat{\bm{d}}_1\right)',\label{eq:P_sim_def}\\
&&\mc{Q}\equiv{q}_{{\rm mag}3}\,,\label{eq:Q_sim_def}
\end{eqnarray}
which will be used to formulate the discrete version of the equations of motion.
Note that $q_{\rm mag1}$ and $q_{\rm mag2}$ are now included in $\bm{\mc{P}}$, appropriately. 
In the next subsection, we incorporate the elasto-magnetic force and torque through $\bm{\mc{P}}$ and ${\mc{Q}}$ into the equations of motion for the centerline position and the twist angle $(\bm{r},\chi_{3})$.

\subsection{Equations of motion used in the discrete simulations}\label{sec:EoM_sim}

In this subsection, we derive the discrete version of the equation of motion for the MRE rod by adding $(\bm{\mc{P}}, \mc{Q})$ into the equation of motion for the centerline position and the twist angle $(\bm{r},\chi_{3})$. 
We start from the magnetic Kirchhoff equations that we derived earlier, with the sole addition of the inertial terms
\begin{eqnarray}
\rho A\frac{\partial {\bm v}}{\partial t}&=&\frac{\partial{\bm F}}{\partial s}+ {\bm p} + \bm{\mc{P}},\label{eq:FeqDy}\\
\frac{\partial {\bm L}}{\partial t} &=& 
\frac{\partial{\bm M}}{\partial s}+ \hat{\bm{d}}_3\times{\bm F} +{\bm q}+ \mc{Q}\hat{\bm{d}}_3\label{eq:KeqDy},
\end{eqnarray}
where $({\bm p}_{\rm mag},{\bm q}_{\rm mag})$ are replaced by $(\bm{\mc{P}},\mc{Q}\hat{\bm{d}}_3)$, as derived in the previous section. The left hand sides of Eq.~(\ref{eq:KeqDy}) represent the inertia terms.
Here, we define the velocity of the centerline $\bm{v}(s,t)={\partial{\bm r}}/{\partial t}$, the angular momentum per unit length ${\bm L} = {I}_a\omega_a\hat{\bm{d}}_a$, and the angular velocity vector ${\bm \omega}(s,t) = \omega_a\hat{\bm{d}}_a$,
with the mass density per unit volume $\rho$ and the principal moments of inertia of the cross section $I_1 = I_2 = I$ and $I_3 = I_1 + I_2 = 2I$~\cite{Goldstein:Book}.
The angular velocity vector $\bm{\omega}$ describes the rotation rates of $\hat{\bm{d}}_a$ in time as
\begin{eqnarray}
\frac{\partial\hat{\bm{d}}_a}{\partial t} = {\bm \omega}\times\hat{\bm{d}}_a\,.
\end{eqnarray}
To ensure that the rod relaxes to the mechanical equilibrium, we include {\it numerical} drag forces and torque acting on the rod-centerline per unit length through ${\bm p} = -\gamma_t{\bm v}$ and ${\bm q}=-\gamma_r\omega_3\hat{\bm{d}}_3$.
We readily derive the dynamic equations of twist around $\hat{\bm{d}}_3$ by taking the inner product between the moment balance equation and $\hat{\bm{d}}_3$. Then, introducing the internal elastic forces ${\bm f}$ and the axial torque ${T}$ per unit length as
\begin{eqnarray}
{\bm f} \equiv \frac{\partial{\bm F}}{\partial s},~~{T} \equiv \frac{\partial{M_3}}{\partial s} + M_2\Omega_1 - M_1\Omega_2,
\end{eqnarray}
we rewrite Eqs.~(\ref{eq:FeqDy}) and (\ref{eq:KeqDy}) as 
\begin{eqnarray}
\rho A\frac{\partial^2 {\bm r}}{\partial t^2} &=& {\bm f} + {\bm p} + \bm{\mc{P}},\\
I_3\frac{\partial^2 \chi_3}{\partial t^2} &=& {T} -\gamma_r\frac{\partial \chi_3}{\partial t} + {\mc Q}
\label{eq:Knodedy},
\end{eqnarray}
respectively. 
We solve Eq. (\ref{eq:Knodedy}) for the discrete versions of the position vector ${\bm r}_i$ and the twist angle $\chi_{3,i}$. 
Choosing the units of mass, length, and force as $m\ell_0$, $\ell_0$ and $EI/\ell_0 ^2$, respectively, we non-dimensionalize Eq.~(\ref{eq:Knodedy}) as
\begin{eqnarray}
\frac{\partial^2 \tilde{{\bm r}}_i}{\partial \tilde{t}^2} &=& \tilde{\bm f}_i + \tilde{\bm p}_i + \tilde{\bm{\mc{P}}}_i,\\
\tilde{I}_3\frac{\partial^2 \chi_{3,i}}{\partial \tilde{t}^2} &=& \tilde{T}_i -\tilde{\gamma}_r\frac{\partial \chi_{3,i}}{\partial \tilde{t}} + \tilde{\mc Q}_i
\label{eq:Kdiscdy},
\end{eqnarray}
where quantities with $(\tilde{\cdot})$ represent the corresponding dimensionless variables.

The numerical scheme for calculating the elastic force ${\bm f}_i$ and axial torque ${T}_i$ at each node $i$ follows that of Refs.~\cite{Chirico:1994hm,Wada:2007dt,Wada:2007kw,Vogel:2010gc,Reichert:2006Thesis}.
In short, we employ the Euler angle representation to describe the configuration of the discrete director frames at each point. 
The Darboux vector components in the discretized model, $\Omega_{a,i}~(a=1,2,3)$, and the corresponding discretized elastic energy can be expressed in terms of the three Euler angles.
The variation of the elastic energy $E_{\rm el}$ is related to the variations of ${\bm r}_i$ and $\chi_{3,i}$ through the variations of the corresponding Euler angles, from which we find the bending and twisting forces acting on each node of the chain~\cite{Chirico:1994hm,Wada:2007dt,Wada:2007kw,Vogel:2010gc,Reichert:2006Thesis}.
All values of the dimensionless parameters are taken from the experiments, which will be presented in Sec.~\ref{sec:Exp}, such that our simulations do not contain any free parameters.
We performed the simulation for a sufficiently long time such that the rod relaxes to the mechanical equilibrium (typically $10^{7}-10^{9}$ discrete time steps), where the number of nodes lies within the range $40\le N \le 120$.
We use the two-step Adams-Bashforth method~\cite{Ref:ABmethod} to numerically integrate the re-scaled dynamical equations in Eq. \eqref{eq:Kdiscdy} with non-dimensional time steps, typically ranging between $10^{-3}-10^{-1}$, to ensure sufficient numerical accuracy.
The director frames at each node are also updated at each time step, and the corresponding Euler angles are calculated for a new configuration.

\section{Experimental fabrication, apparatus, and protocols}
\label{sec:Exp}


We perform three different sets of experiments to validate our theory presented in Sec. \ref{sec:Theory}: (i) a naturally straight rod under a constant field, (ii) a helical rod under a constant field, and (iii) a helical rod under a constant gradient field. We chose these three configurations because previous works~\cite{Wang:1986kl,Ciambella:2020je,Wang:2020du,Yan:2020prep} were limited to the case of straight beams, without twist deformations. For case (i), we still consider a straight rod but also include twist deformation. Through case (ii), we can individually validate the elasto-magnetic torque $\bm{q}_{\rm mag}$. In case (iii), both $\bm{q}_{\rm mag}$ and $\bm{p}_{\rm mag}$ can be validated. 
In this section, we present the details of our experiments, whose results will be provided in Sec. \ref{sec:3D}. 
To perform the experiments, in Sec.~\ref{sec:fab}, we detail the procedure that we developed to manufacture our MRE rods. In Sec.~\ref{sec:apparatus}, we present the design of the magnetic coils used to generate the magnetic field. The detailed protocols for the cases (i)-(iii) follow in Sec.~\ref{sec:subproto}.

\subsection{Fabrication of hard magnetic rods}
\label{sec:fab}

Our rods were made of a magnetorheological elastomer (MRE), a composite of hard-magnetic NdPrFeB particles and vinylpolysiloxane (VPS) polymer (Young's modulus $E_{\rm vps} = 1.16\pm0.03~{\rm MPa}$ and mass density $\rho_{\rm vps} = 1.17\pm0.26~{\rm g/cm^3}$). The fabrication of the MRE rods involved the following steps. First, the non-magnetized NdPrFeB particles (average size of 5$\mu$m, mass density $\rho_{\rm mag} = 7.61~{\rm g/cm^3}$ MQFP-15-7-20065-089, Magnequench) of weight $m_{\rm mag}$ were mixed with the VPS base liquid solution (Elite Double 32, Zhermack) of weight $m_{\rm base}$ using a centrifugal mixer (ARE-250, Thinky Corporation). Secondly, the VPS catalyst (weight $m_{\rm cat} = m_{\rm base}$) was added to the mixed solution with a ratio of 1:1 in weight to the VPS base. The solution of total weight $m_{\rm tot} = m_{\rm mag} + m_{\rm base} + m_{\rm cat}$ was then mixed using the centrifugal mixer for 40 s at 2000 rpm (clockwise), and another 20 s at 2200 rpm (counterclockwise). We further degassed the solution in a vacuum chamber (absolute pressure below 8 mbar) to remove any air bubbles that could otherwise compromise the homogeneity of the MRE.
The final solution was injected into the molds (detailed in the next paragraph) using a syringe to cast either straight or helical rods. 
We varied the mass concentration ratio for NdPrFeB particles in the range $c = m_{\rm mag}/m_{\rm tot} = 10-30\%$. 

\begin{figure}[!ht]
    \centering
    \includegraphics[width=1.0\textwidth]{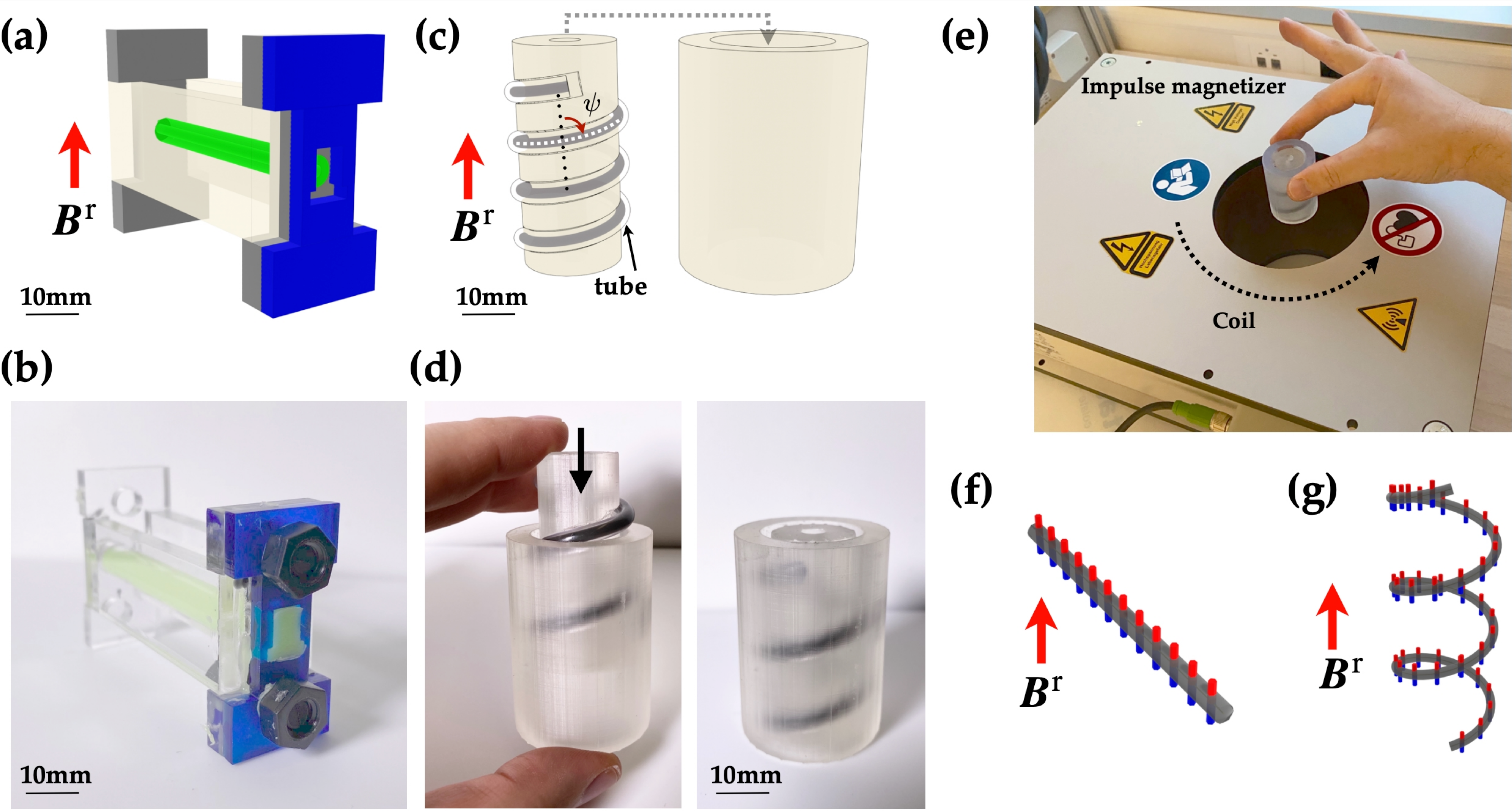}
    \caption{Fabrication procedure to manufacture straight and helical rods made out of a MRE.  
    (a) ~Schematic diagram of the mold for a straight rod. A straight acrylic tube (green) is threaded into two parallel supporting plates. 
    (b) ~Photograph of the mold used to produce a straight rod. 
    (c) ~Schematic diagram of the mold for a helical rod. A rigid cylinder with a helical groove (inner mold) is 3D printed. 
    (d) ~Photograph of the mold for a helical rods. 
    (e) The mold with the cured elastomer is placed into an impulse magnetizer such that ${\bm B}^{\rm r}$ in (a) and (c) is aligned with the axis of the coil. 
    Schematics of a magnetized (f) straight and (g) helical rod. The local magnetization vectors are depicted by the embedded ``magnets," where the red and blue ends correspond to the north and south poles, respectively. 
    }
    \label{fig:Mold}
\end{figure}

The schematic and a photograph of a mold used to fabricate a straight rod are shown in Fig.~\ref{fig:Mold} (a) and (b), respectively. 
We injected the mixed solution into a straight acrylic tube (Plexiglas XT tubes incolore 0A070, R\"ohm, Switzerland, inner diameter $2$ or $4$~mm) threaded into two coaxial holes in acrylic plates. One of the inlets of the tube was designed to have a convex shape pointing to the direction of the residual flux density.
The convex inlet is utilized to align the magnetization vector ${\bm B}^{\rm r}$ precisely.

To fabricate helical rods, we prepared a rigid cylinder of diameter $20$~mm with a helical half-piped groove (diameter $4.5$ mm) using a 3D printer (Printer: Form 2, Formlabs, Material: Clear Resin (RS-F2-GPCL-04)). 
The schematic and the photograph of the mold with a helical groove are shown in Fig.~\ref{fig:Mold} (c) and (d), respectively. 
A flexible tube (diameter $4$ mm, Misumi) is placed along the groove, injecting the mixed solution into it.
The inner mold and the tube were then positioned inside a 3D printed (outer) cylinder (inner diameter $12$ mm and outer diameter $16$ mm). 
The central axis of the helical groove and that of the outer cylinder were set co-axially. 
The 3D printed molds allowed us to control the pitch angle $\psi$ of a helical MRE rod up to $\psi\lesssim1.49$ rad. 
For larger values of $\psi$, we prepared the acrylic cylinder (without a groove) and tightly spooled a flexible tube around it with a pitch angle of $\psi = 1.51$ rad.

After injection of the mixed solution into either the straight or helical molds, the curing of the polymer mixture occurred in approximately $20$ min, at room temperature. 
The (projected) radius of the helical centerline was chosen to be $R = 10~{\rm mm}$ to minimize the deformation of the tube, allowing us to have a cross-section of the helix close to circular, which could otherwise deviate from the circular shape for tighter helices due to the Brazier instability~\cite{Brazier:1927du}.

Before demolding, the rods were magnetized permanently by saturating the NdPrFeB particles in the MRE using an applied uniaxial magnetic field (4.4 or 2.5~T) generated by an impulse magnetizer (IM-K-010020-A Magnet-Physik Dr. Steingroever GmbH), as shown in Fig.~\ref{fig:Mold} (e). The directions of ${\bm B}^{\rm r}$ for either a straight or helical rods were set normal to the cross section or parallel to the central axis of a helix, respectively. Schematic diagrams for the magnetized straight and helical rods are shown in Figs.~\ref{fig:Mold} (f) and (g), respectively. The fabricated MRE rods possessed the residual flux densities of $|{\bm B}^{\rm r}| = 0.90c_v~{\rm T}$ (Secs.~\ref{sec:Twist} and \ref{sec:Helix}) and $|{\bm B}^{\rm r}| = 0.86c_v~{\rm T}$ (Sec.~\ref{sec:HelixBuckling}). The Young's modulus was $E = E_{\rm vps}/(1-c_v ^{1/2})$ with the volume concentration of the particles equal to $c_v = (1 + (\rho_{\rm mag}/\rho_{vps})(m_{\rm tot}/m_{\rm mag} - 1))^{-1}$~\cite{Counto:1964bh}. After magnetization, the rods were demolded and cut to the desired length $L$ (within an experimental uncertainty of $\pm1$~mm).

\subsection{Experimental apparatus}
\label{sec:apparatus}

During the experimental tests, the samples are loaded magnetically by placing them in between a set of two coaxial coils, which induce a steady axial symmetric magnetic flux ${\bm B}^{\rm a}(r,z)$~\cite{Yan:2020prep}. 
Each coil is manufactured by winding an aluminum circular spool with an insulated magnet wire (Repelec Moteurs S.A.). The magnet wire (enameled wire, Isomet AG) has a circular cross-section of diameter $1.32$ mm for the copper core and thickness $0.102$ mm for the outer insulation layer. The final dimensions of the coil are $86$ mm in inner diameter, $152$ mm in outer diameter, and $43$ mm ($33$ mm for wires and $10$ mm for the spool) in height; the mean radius of the coil $\mc{R}$ is $\mc{R} = 59.5$~mm. 
The coils are powered by a DC power supply providing a maximum current to power ratio of $25$ A/ 1.5kW (EA-PSI 9200-25T, EA-Elektro-Automatik GmbH).

The two identical coils are set co-axially along $\hat{e}_z$ and their current can flow in either the same or opposite directions. 
To realize a {\it constant (homogeneous) field}, the center-to-center axial distance is set to be $59.5$~mm$(=\mc{R})$, and the current in the coils is set to flow in the same direction; this is known as the Helmholtz coil configuration. 
In the central region between the coils, the field generated by the Helmholtz coil is 
\begin{eqnarray}
{\bm B}^{\rm a} = B^{\rm a} \hat{e}_z.
\label{eq:Helmholtz}
\end{eqnarray}
In contrast, to realize the {\it constant gradient field}, the center-to-center axial distance is set to be $103$ mm $(=\sqrt{3}\mc{R})$, and the current in the coils is set to
flow in opposite directions. This configuration is known as the Maxwell coil, where the magnetic field induced near the coil is 
\begin{eqnarray}
{\bm B}^{\rm a}(r,z) = -\frac{b}{2}r\hat{e}_r + b z \hat{e}_z.
\label{eq:Maxwell}
\end{eqnarray}
In this Maxwell coil configuration, the gradient of the magnetic field is $b = (\partial_z{\bm B}^{\rm a})\cdot\hat{e}_z$, i.e. constant. From the Gauss law of magnetism, $\nabla\cdot\bm{B}^{\rm a} = 0$, there are gradient components in both the radial, $\bm{e}_r$, and the $\bm{e}_z$ directions.

\begin{figure}[!ht]
    \centering
    \includegraphics[width=1.0\textwidth]{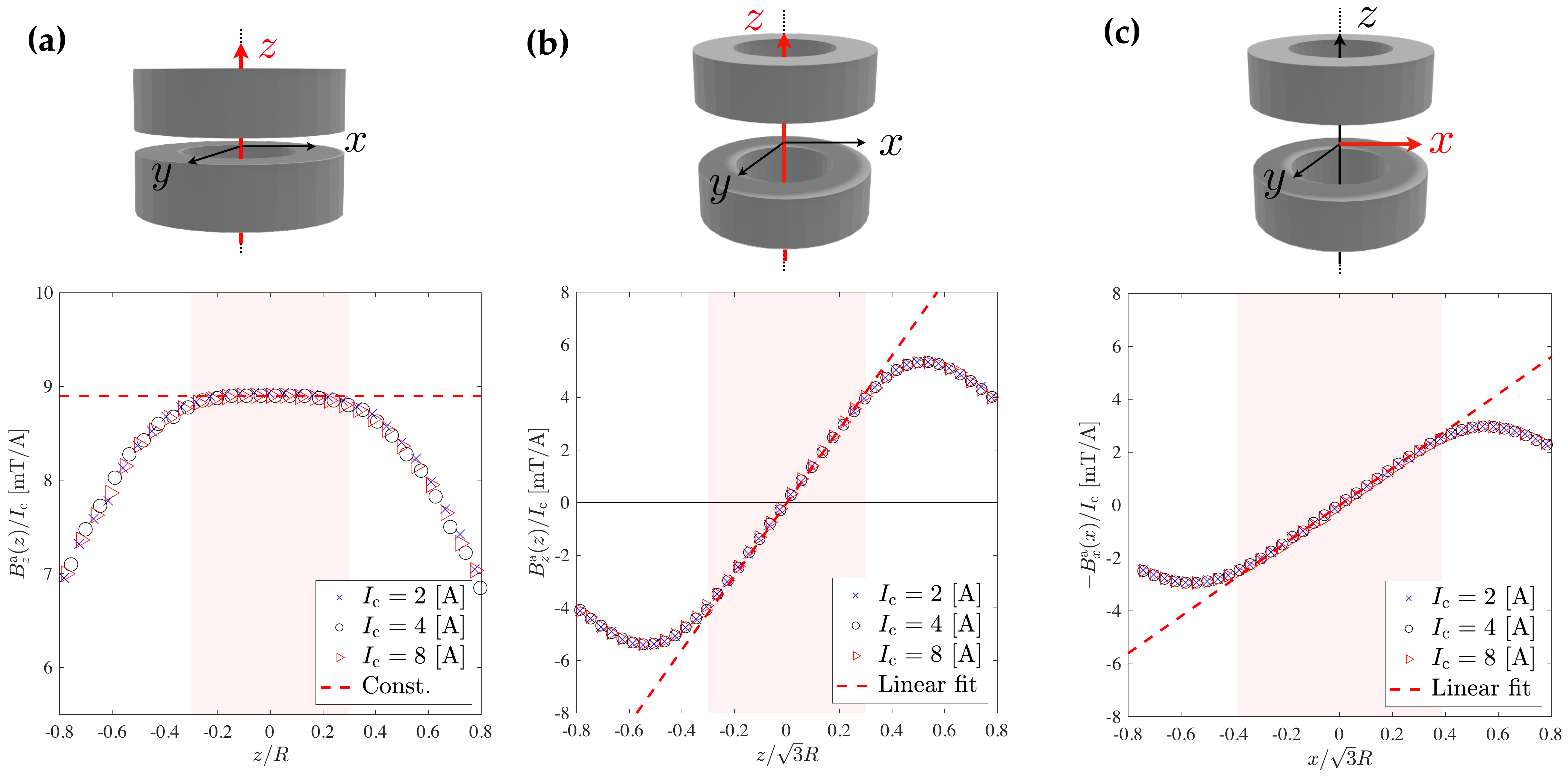}
    \caption{
    Experimental characterization of the field generated by the magnetic coils in the (a)~Helmholtz (constant field) and (b, c)~ Maxwell (constant gradient field) configurations. 
    (a) Profile of the (rescaled) applied field along $x=y=0$. Profile of the applied field along (b) $x=y=0$ and (c) $y=z=0$. The dashed lines in the plot correspond to the (a)~constant and (b) (c)~linear fits within the shaded regions where the experiments were performed.}
    \label{fig:Coil}
\end{figure}

We characterized the relationship between the flux density and the current in the coils, $I_c$, by using a Teslameter (FH 55, Magnet-Physik Dr. Steingroever GmbH) and the results are plotted in Fig.~\ref{fig:Coil}. 
The fields described by Eqs.~(\ref{eq:Helmholtz}) and (\ref{eq:Maxwell}) are realized in the region of $z,r\lesssim 0.6\mc{R}$. 
For the Helmholtz configuration, the constant field is induced near the center, for several values of $I_c$. In Fig.~\ref{fig:Coil} (a), we plot the profiles of the measured field $\bm{B}^{\rm a}$ divided by $I_c$, which collapse onto a single curve. This collapse confirms that, $\bm{B}^{\rm a}$ is proportional to $I_a$. By fitting the constant $B^{\rm a}$ near the center, we find $\bm{B}^{\rm a}$ and $I_a$ as $B_a/I_c = 8.9~{\rm mT/A}$.
For the Maxwell configuration, a constant gradient field is generated near the center of the coils (Fig.~\ref{fig:Coil} (b) and (c)). 
We measured the profiles of the induced field both along with $\bm{e}_z$ and $\bm{e}_r$ for different values of $I_c$. The data again collapse on a single curve as $\bm{B}^{\rm a}/I_c$, which implies that the gradient of the field $b$ is linearly proportional to $I_c$. 
By fitting the field along $\bm{e}_z$, we find $b/I_c = 0.133~{\rm T /(A\cdot m)}$. From Eq.~(\ref{eq:Maxwell}), the gradient along $\bm{e}_r$ is half of that along $\bm{e}_z$. In Fig.~\ref{fig:Coil} (c), we plot the slope of $b/2$ as the dashed line, which is in an excellent agreement with the measurement. 
This experimental characterization confirms that our coils accurately produce the fields according to Eqs.~(\ref{eq:Helmholtz}) and (\ref{eq:Maxwell}) when set in either the Helmholtz or the Maxwell configurations, respectively.

\subsection{Experimental protocols}
\label{sec:subproto}

\begin{figure}[!ht]
    \centering
    \includegraphics[width=1.0\textwidth]{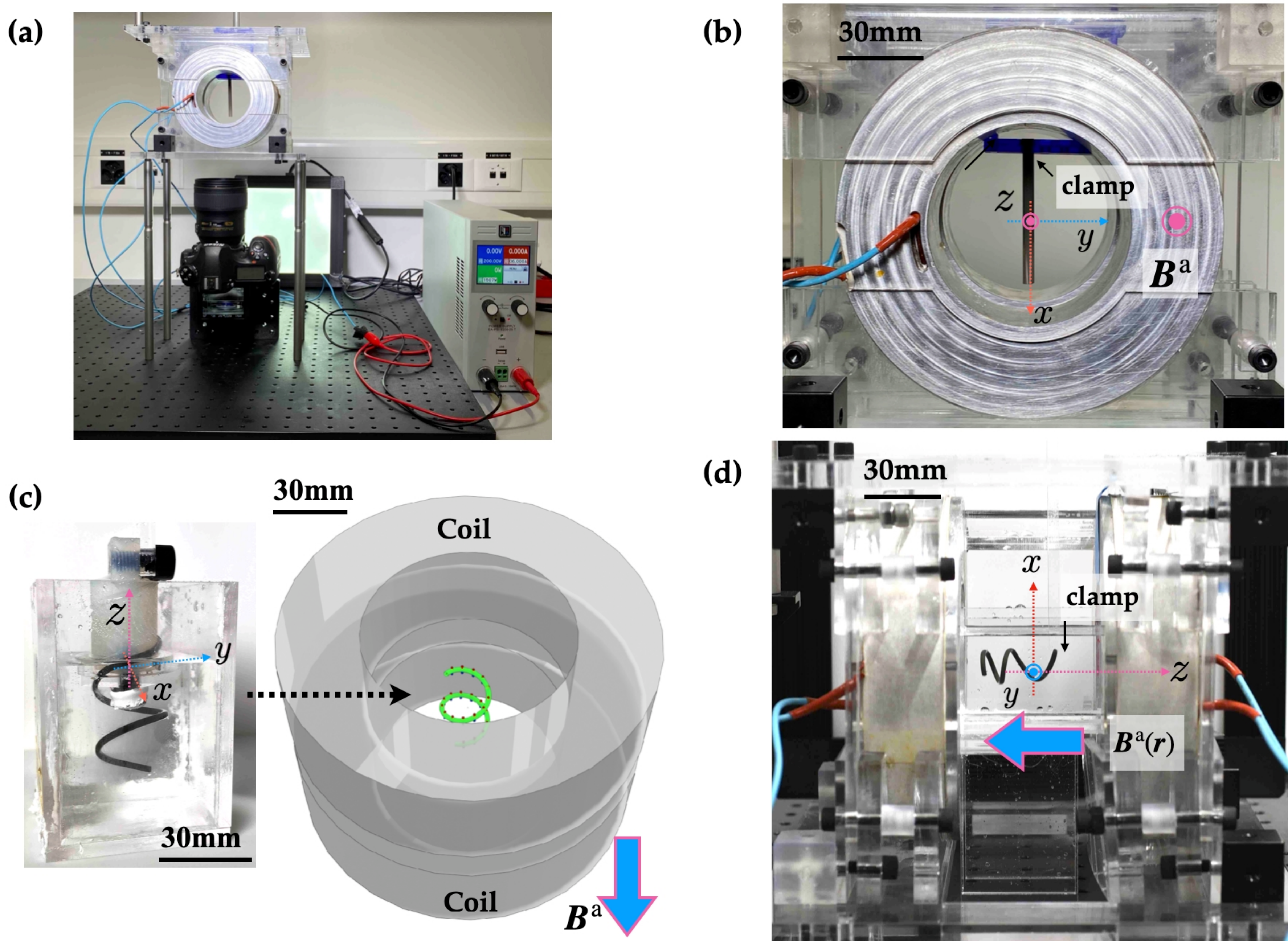}
    \caption{
    (a) Photograph of the full experimental apparatus. A single power supply generates the electric current provided to the two coils. A digital camera set underneath the coils captures the deformation of the rod. 
    (b) Experimental setup used for the twist instability of a straight rod (Sec.~\ref{sec:Twist}). The straight rod magnetized along $-\hat{e}_z$ is clamped along $\hat{e}_x$. The applied field ${\bm B}^{\rm a}$ is set as shown in Eq.~(\ref{eq:Helmholtz}). 
    (c) The schematic of the experimental setup for a helix under constant field (right panel, See Sec.~\ref{sec:HelixBuckling}), with the uniformly helical magnetic rod (left panel). The helix magnetized along $\hat{e}_z$ is clamped such that its central axis is parallel to that of the coils. The magnetic helix is immersed in a glycerol bath.
    (d) Photograph of a magnetic helical rod (in the glycerol bath) under constant gradient field (Sec.~\ref{sec:Helix}). 
    }
    \label{fig:Protocol}
\end{figure}

Having described the experimental apparatus above, we proceed by presenting the protocols that we followed during the experiments. We placed the samples (prepared through the procedure detailed in Sec.~\ref{sec:fab}) between the coils (whose field was characterized in the previous section). 
First, we will study the deformation of a straight or curved rod under the constant field (Helmholtz coil) followed by the study of the deformation of a helix under the constant gradient field (Maxwell coil), as shown in Fig.~\ref{fig:Protocol} (a)-(d). 
A photograph of our apparatus is shown in Fig.~\ref{fig:Protocol} (a). 
All the experimental results for the corresponding setups will be presented in Sec.~\ref{sec:3D}.
The protocols for each experiment are detailed next.

In Sec.~\ref{sec:Twist}, we will present results on the deformation of a straight rod under constant field. 
For these experiments, our MRE rods were magnetized along $-\hat{\bm{e}}_z$, and a constant magnetic field was applied along $\hat{\bm{e}}_z$; see Fig.~\ref{fig:Protocol} (b). 
An acrylic clamp, engraved to have the convex shape at one of the ends of the rod, hanged the rod from above. This clamp was also designed such that its width matched the gap between the coils ($13.5~{\rm mm}$). 
We varied the total length of the rod in the ranges $L = 40-57$~mm, the mass concentration ratio of NdPrFeB particles is $c = 10,~20,~30\%$, and the rod diameter was $d = 3,5$~mm. 
We measured the twist deformation of the rod (detailed in the next section) using a digital camera, placed underneath the setup and directed upwards. The measured twisting angles were averaged over 4 or 5 samples. 

In Sec.~\ref{sec:HelixBuckling}, we will study the deformation of a helical rod under the constant field.
For these experiments, the helix was clamped vertically by a 3D-printed rigid cylinder such that its central axis was parallel to the coil axis ($\hat{\bm{e}}_z$). 
The constant magnetic field (Eq.~(\ref{eq:Helmholtz})) was applied (anti-parallel to $\bm{B}^{\rm r}$) to the helical rod magnetized along $\hat{\bm{e}}_z$, as shown in Fig.~\ref{fig:Protocol} (c). The experiments were performed in a glycerol bath ($85$\%, $1.23$ ${\rm g}/$cm$^3$, Sigma-Aldrich), which density-matched the rods, thereby minimizing the effect of gravity. We performed experiments for the MRE rod of diameter $2$~mm, pitch angle $\psi=1.28$~rad, and radius of curvature $R = 10$~mm. The total length $L$ and the concentration ratio $c$ were varied in the ranges $L = 40-120$~mm and $c = 10, 20\%$, respectively. 

For the results in Sec.~\ref{sec:Helix} and \ref{sec:Limitation}, we applied a constant gradient field (according to Eq.~(\ref{eq:Maxwell})) to a helical rod, whose central axis in the reference configuration was chosen to be aligned with the axis of the coils ($z$-axis, perpendicular to gravity). As shown in Fig.~\ref{fig:Protocol} (d), we clamped the helix and perform the experiment in a glycerol bath. During the experiments, we imaged the deformed shape of the helix using a digital camera.  
In Sec.~\ref{sec:Helix}, we measured the displacement of the free-end as a function of the applied field. The total length $L$ and the concentration ratio $c$ were varied as $L = 74,81,103$~mm and $c=10,20\%$, respectively, while the diameter $d = 2$~mm, the radius of natural curvature $R = 10$~mm and the pitch angle $\psi = 1.51$~rad were fixed. In Sec.~\ref{sec:Limitation}, the deformation of the magnetic helix were studied, varying the pitch angle $\psi$ and total length $L$ in the range of $\psi = 1.26-1.49$~rad and $L = 65-140$~mm, while $d = 2$~mm, $R = 10$~mm, and $c = 20\%$ were kept fixed throughout Sec.~\ref{sec:Limitation}. 

\section{Validation of the theory of hard magnetic rods vs experiments}
\label{sec:3D}

In this section, we will use precision experiments on specific test configurations to perform a detailed validation of the theoretical framework introduced in Sec.~\ref{sec:vw} for hard MRE rods. 
In Sec.~\ref{sec:Twist} and \ref{sec:HelixBuckling}, the twist instabilities for a straight rod and a helix are investigated under a constant field. Given that $\bm{p}_{\rm mag} = 0$ under the constant field, we can show that the elasto-magnetic torque $\bm{q}_{\rm mag}$ in both naturally straight and curved configurations captures the corresponding experimental results in excellent agreement. Then, in Sec.~\ref{sec:Helix}, we study the deformation of a helix under a constant gradient field to test the validity of both $\bm{p}_{\rm mag}$ and $\bm{q}_{\rm mag}$.

\subsection{Twist instability of a straight magnetic rod under magnetic loading}
\label{sec:Twist}

\begin{figure}[!ht]
    \centering
    \includegraphics[width=1.0\textwidth]{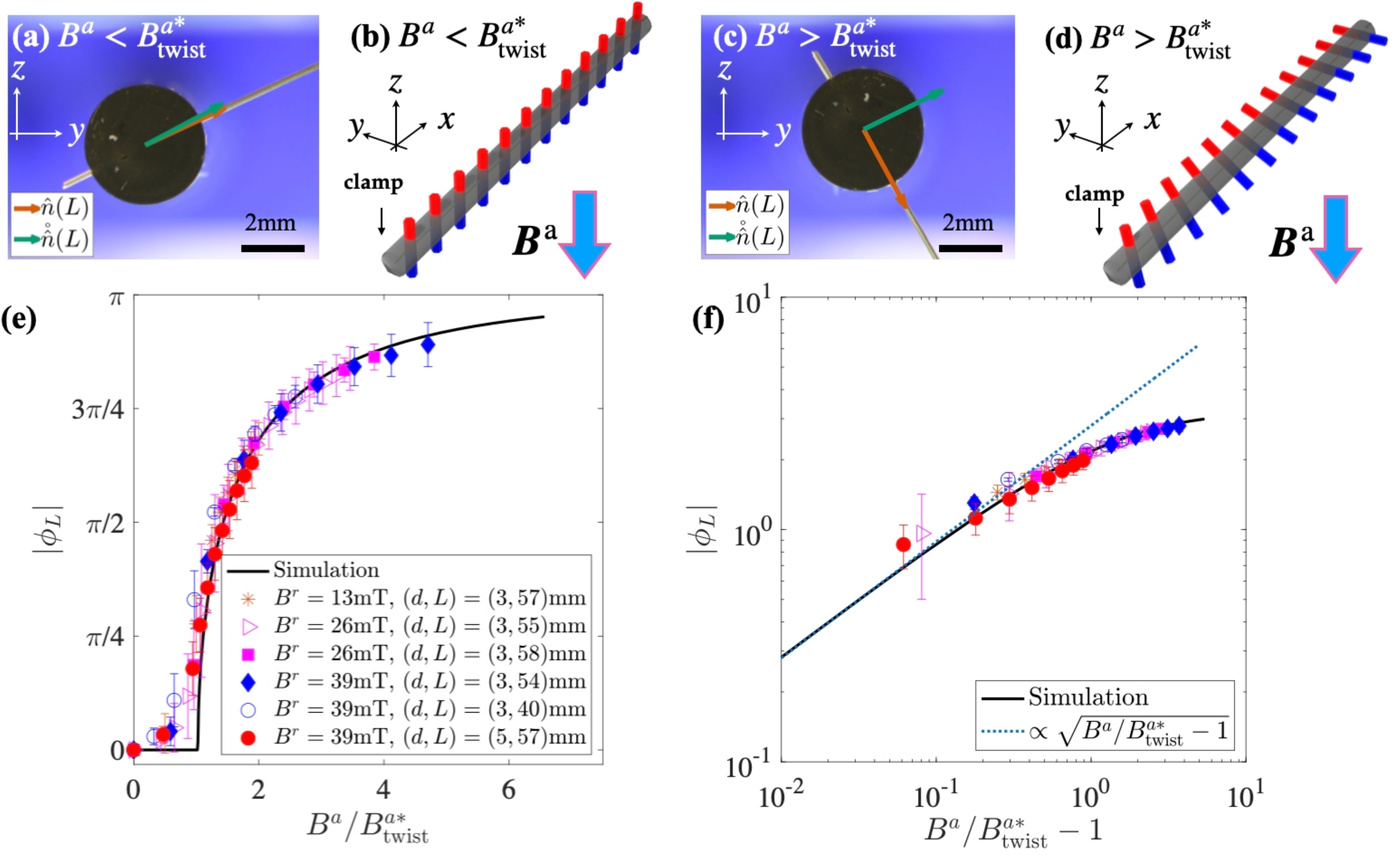}
    \caption{Twist instability of straight MRE rods. (a) Experimental photograph of the cross section at $s = L$ with $B^{\rm a} < B^{\rm a*}_{\rm twist}$. The symbols $\hat{n}(L)$ and $\mr{\hat{n}}(L)$ represent the direction of the nitinol rod in the deformed and reference configuration, respectively. The corresponding simulation snapshot is shown in (b).
    (c) Experimental photograph of the cross section at $s = L$ with $B^{\rm a} > B^{\rm a*}_{\rm twist}$. The corresponding simulation snapshot is shown in (d). (e) $\phi_L$ as a function of $B^{\rm a}/B^{\rm a*}_{\rm twist}$. The data points are experimental results and the solid line is the simulation result. (f) Logarithmic plot of $\phi_L$ as a function of $B^{\rm a}/B^{\rm a*}_{\rm twist} - 1$. The legend is the same as (e). 
    The dashed line is the fitting of the power law with the exponent $1/2$, Eq.~(\ref{eq:twist_scaling}), obtained through scaling arguments.}
    \label{fig:Twist}
\end{figure}

Consider a straight rod aligned with $\hat{\bm{e}}_x$ (clamped at $s = 0$), that is magnetized normal to the tangent, i.e., ${\bm B}^{\rm r} = B^r\hat{\bm{e}}_z$. 
We apply the constant flux density as ${\bm B}^{\rm a} = -B^a\hat{\bm{e}}_z$ (see~Fig.~\ref{fig:Twist} (a)-(d)). 
Figures~\ref{fig:Twist} (a) and (c) are photographs from the experiments of the rod cross-sections at their free-ends, parallel with $y$-$z$ planes. Three-dimensional representations of the straight MRE rods obtained from the simulations are shown in Figs.~\ref{fig:Twist} (b) and (d). 
The rod remains straight when the magnitude of the applied field $B^{\rm a}$ is sufficiently small enough (Fig.~\ref{fig:Twist} (a) and (b)), whereas the rod twists, with the centerline still straight, above the critical flux density $B^{a*}_{\rm twist}$ (Fig.~\ref{fig:Twist} (c) and (d)). 
In the experiments, we measured the twist angle $\phi_L$ at $s = L$ by tracking the rotation angle of a nitinol rod embedded near the free-end perpendicular to the centerline (as shown by $\hat{n}(L)$ in Figs.~\ref{fig:Twist} (a) and (c)). 
In the simulation, we also measured $\phi_L$ by computing the rotation angle for one of the basis vectors of the Cosserat frame $\hat{\bm{d}}_1$.
In Fig.~\ref{fig:Twist} (e), we superpose the experimental and numerical results of $\phi_L$ as a function of $B^{\rm a}/B^{\rm a*}_{\rm twist}$. The error bars in the experimental data correspond to the uncertainty in the setting of the angle of the clamp. Experimental and numerical results are in excellent agreement with each other across the full parameter range in $B^{\rm a}/B^{\rm a*}_{\rm twist}$.

We rationalize the experimental and numerical results using the magnetic Kirchhoff equations that we derived in Sec.~\ref{sec:vw} by the perturbation against the twist instability. 
Following Ref.~\cite{Goriely:1997hw}, we expand the Cosserat frame with respect to the straight configuration $(\hat{\bm{d}}_1 ^{(0)},\hat{\bm{d}}_2 ^{(0)},\hat{\bm{d}}_3 ^{(0)}) = (\hat{\bm{e}}_y,\hat{\bm{e}}_z,\hat{\bm{e}}_x)$ as
\begin{eqnarray}
\hat{\bm{d}}_a = (\delta_{ab} + \varepsilon_{abc}\alpha_c)\hat{\bm{d}}_b ^{(0)},
\label{eq:da_perturb}
\end{eqnarray}
where $\varepsilon_{abc}\alpha_c$ is the asymmetric tensor and $\alpha_a=\alpha_a(s)~(a=1,2,3)$ is the small (angular) perturbation parameter $\alpha_a\ll1$. 
The coefficient $\varepsilon_{abc}\alpha_c$ enforces the orthogonality of $\hat{\bm{d}}_a~(a=1,2,3)$ even up to the 1st order of the perturbation. 
Plugging the perturbed form of $\hat{\bm{d}}_a$, Eq.~(\ref{eq:da_perturb}), into the kinematic equation, Eq.~(\ref{eq:kinematic}),~(see~\ref{sec:App1} for details), we find 
\begin{eqnarray}
\alpha_a' = \Omega_a\,,
\label{eq:alpha_omega}
\end{eqnarray}
from which we conclude that $\alpha_a$ represents the infinitesimal rotation angle around $\hat{\bm{d}}_a$. 
Given that the rod is clamped at $s = 0$ and free at $s = L$, the corresponding boundary conditions are $\alpha_a(0) = 0$ and $\alpha_a '(L) = \Omega(L) = 0$. 
Furthermore, ${\bm p}_{\rm mag} = 0$ and thus ${\bm F}(s) = 0$. 
Substituting Eq.~(\ref{eq:da_perturb}) into Eq.~(\ref{eq:Meq}) and linearizing the result for $\alpha_a$ yields
\begin{eqnarray}
&&EI\alpha_1 '' = - \frac{AB^{\rm r}B^{\rm a}}{\mu_0}\alpha_1,\label{eq:TwistPerturb1}\\
&&EI\alpha_2'' = 0,\label{eq:TwistPerturb2}\\
&&GJ\alpha_3 '' = - \frac{AB^{\rm r}B^{\rm a}}{\mu_0}\alpha_3\label{eq:TwistPerturb3}.
\end{eqnarray}
Equations (\ref{eq:TwistPerturb1})-(\ref{eq:TwistPerturb3}) are the governing equations for the infinitesimal rotation angles $\alpha_a$ along the rod.
To determine the conditions for the existence of nontrivial solutions $(\alpha_a(s)\ne0)$ for Eqs.~(\ref{eq:TwistPerturb1})-(\ref{eq:TwistPerturb3}), we will substitute $\alpha_a \propto\sin(\pi s/2L)~(a=1,2,3)$, which satisfies the boundary conditions $\alpha_a(0) = \alpha_a'(L) = 0$, into Eqs.~(\ref{eq:TwistPerturb1})-(\ref{eq:TwistPerturb3}). 
Equation (\ref{eq:TwistPerturb2}) gives us the null amplitude $\alpha_2(s) = 0$, while Equations (\ref{eq:TwistPerturb1}) and (\ref{eq:TwistPerturb3}) are satisfied if ${AB^{\rm r}B^{\rm a}}/{\mu_0} = EI(\pi/2L)^2$ or ${AB^{\rm r}B^{\rm a}}/{\mu_0} = GJ(\pi/2L)^2$ hold, respectively. 
The smaller value of $B^{\rm a}$ derived here corresponds to the critical value of $B^{\rm a}$ for the twist instability, $B^{\rm a*}_{\rm twist}$. Since $GJ/EI = 1/(1+\nu) < 1$, the eigenvalue of Equation (\ref{eq:TwistPerturb3}) corresponds to the critical flux density, above which the rod twists:
Hence, we find
\begin{eqnarray}
B^{\rm a*}_{\rm twist} = \left(\frac{\pi}{2}\right)^2\frac{\mu_0 GJ}{AB^{\rm r}L^2}.
\label{eq:Btwist}
\end{eqnarray}
Notice that $EI$ in Eq.~(\ref{eq:Bbend}) is now replaced by $GJ$ in Eq.~(\ref{eq:Btwist}).
Given that $\alpha_3 = \alpha_3(s)$ is the profile of the (infinitesimal) twisting angle (from Eq.~(\ref{eq:alpha_omega})), $\alpha_3(L)$ corresponds to the twist angle of the free-end measured in our experiment and $\phi_L=\alpha_3(L)$ in the limit of $\phi_L\ll1$.
By analogy with the equation of clamped-free elastica~\cite{Audoly:2010Book}~(replacing $\alpha_3(s)$ with the bending angle in elastica), Equation~(\ref{eq:TwistPerturb3}) is now mathematically equivalent to the amplitude equation for the pitch-fork bifurcation.
In the neighborhood of $B^{\rm a}/B^{\rm a*}_{\rm twist}\sim 1$, we thus anticipate that $\phi_L$ obeys the following scaling law, where the twist angle at the tip $\phi_L$ evolves as 
\begin{eqnarray}
\phi_L\propto\sqrt{\frac{B^{\rm a}}{B^{\rm a*}_{\rm twist}}- 1}.
\label{eq:twist_scaling}
\end{eqnarray}

We find an excellent agreement between experiments and the theoretical description developed above $\phi_L$, although the transition is not sharp due to the inevitable imperfection of the system below the critical field $B^{\rm a}/B^{\rm a*}_{\rm twist}<1$~\cite{Strogatz:Book}.
Furthermore, from the logarithmic plot in Fig.~\ref{fig:Twist} (f), we confirm that $|\phi_L|$ is consistent with the predicted scaling law Eq.~(\ref{eq:twist_scaling}), in the region of $|B^{\rm a}/B^{\rm a*}_{\rm twist}-1|\ll1$.
The above results demonstrate that our magnetic Kirchhoff equations correctly predict the twist instability of straight rods.
In particular, we validated the elasto-magnetic torque $\bm{q}_{\rm mag}$ in the case of straight rods. Combining the fact that our framework reproduces the previous results on the planar deformation (Sec.~\ref{sec:Reduc2D})~\cite{Lum:2016fc,Ciambella:2020je,Wang:2020du,Yan:2020prep} and the results presented in this subsection, the magnetic Kirchhoff equation (\ref{eq:Feq}) and (\ref{eq:Meq}) correctly predict the large deformation of naturally straight hard MRE rods. Next, we focus on the deformation of hard MRE rods that are naturally curved .

\subsection{Buckling of a magnetic helix under constant field}
\label{sec:HelixBuckling}

We proceed by considering a naturally curved and twisted rod, i.e., a helix, which undergoes structural (buckling) instability of a helix at a critical field $B^{\rm a*}$, but with an inevitable twist-bend coupling. 
We will show that our simulations and theoretical analyses rationalize our experimental results with excellent quantitative agreement.

\begin{figure}[!ht]
    \centering
    \includegraphics[width=1.0\textwidth]{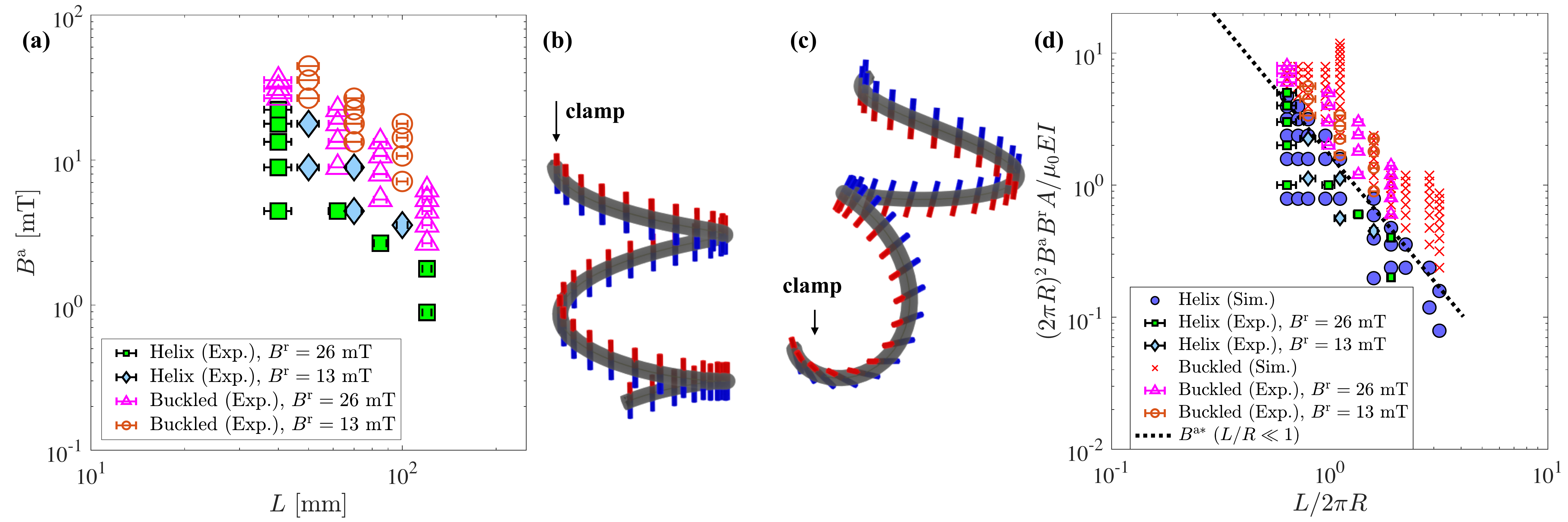}
    \caption{Buckling of a magnetic helix. (a) Phase diagram of the buckling instability of a magnetic helix in experiments. Filled and empty data points are uniform or buckled helix states, respectively. Buckling of a magnetic helix under constant field. In the previous section. The dashed line is the analytic prediction in the limit of $L/R\ll1$. The simulation snapshots for $L/R = 12$ correspond to $B^a / B^{a*} = 0.25$ (b) and $B^a / B^{a*} = 2.5$ (c). (d) Rescaled phase diagram of our experimental results. The simulation results and asymptotic solutions are superposed with $\psi = 1.28~{\rm rad}$.}
    \label{fig:HelixBuckling}
\end{figure}

We fabricate a hard magnetic helix with the pitch angle $\psi = 1.28~{\rm rad}$, which is clamped such that the central axis is aligned along the coil axis. 
We apply the magnetic field ${\bm B}^{\rm a} = -B^{\rm a}\hat{\bm{e}}_z$ on a helical rod with lengths in the range of $40\leq L~[{\rm mm}]\leq120$. 
These helices are magnetized along their central axis as ${\bm B}^{\rm r} = B^{\rm r}\hat{\bm{e}}_z$. Below the critical field $B^{\rm a} < B^{\rm a*}$, the rod remains uniformly helical, as shown in the filled data points in Fig.~\ref{fig:HelixBuckling} (a), buckling occurs when $B^{\rm a} > B^{\rm a*}$, as depicted by the open symbols. We regarded the rod as helical if the tip moves by less than the diameter $d$ in the experiments.

We now specialize the magnetic Kirchhoff equations Eqs.~(\ref{eq:Feq}) and (\ref{eq:Meq}) for a magnetic helix to rationalize the experimental observations. The critical flux density for buckling of the magnetic helix $B^{\rm a*}$ can be derived via rigorous analysis of Eqs.~(\ref{eq:Feq}) and (\ref{eq:Meq}). Here, to simplify the analysis and emphasize the importance of a twist-bend coupling in the system, we will compute the critical field strength $B^{a*}$ in the limit of small curvature $KL\ll1$ ($K$ is total curvature for a helix defined later) but with $\psi<\pi/2$, which indeed provides a good prediction for our observations as we detail later.

Similarly to Sec.~\ref{sec:Twist}, we expand the Cosserat frame with respect to the uniform helix as we did in Eqs.~(\ref{eq:da_perturb}) and (\ref{eq:alpha_omega}). The Darboux vector in the uniform helix is ${\bm \Omega} = \kappa{\hat{\bm{d}}}_2 + \tau{\hat{\bm{d}}}_3$ with the natural curvature $\kappa$ and twist $\tau$. 
By integrating the kinematic relation Eq.~(\ref{eq:kinematic}), $\hat{\bm{d}}_a' = {\bm \Omega}\times\hat{\bm{d}}_a$, the Cosserat frame $\hat{\bm{d}}_a~(a=1,2,3)$ for a helix in the reference configuration is computed as
\begin{eqnarray}
    \hat{\bm{d}}_1^{(0)}(s) &=& \cos(K s)\hat{\bm{e}}_x - \sin(K s) \hat{\bm{e}}_y\label{eq:d1_0},\\
    \hat{\bm{d}}_2^{(0)}(s) &=& -\cos\psi\left(\sin(Ks)\hat{\bm{e}}_x + \cos(Ks)\hat{\bm{e}}_y\right) - \sin\psi\hat{\bm{e}}_z\label{eq:d2_0},\\
    \hat{\bm{d}}_3^{(0)}(s) &=& \sin\psi\left(\sin(Ks)\hat{\bm{e}}_x + \cos(Ks)\hat{\bm{e}}_y\right) - \cos\psi\hat{\bm{e}}_z\label{eq:d3_0}.
\end{eqnarray}
We choose the central axis to be parallel to $\hat{z}$ and define the pitch angle $\psi$ as $\kappa = K\sin\psi$, $\tau = \kappa\cos\psi$, and $K \equiv\sqrt{\kappa^2 + \tau^2}$. 
The radius of curvature on the $x-y$ plane $R$ and $\psi$ are related as $\kappa = \sin^2\psi/R$ and $\tau = \sin2\psi/2R$ (see~\ref{sec:App2} for the detailed derivation). 

Plugging Eqs.~(\ref{eq:d1_0})-(\ref{eq:d3_0}) into the magnetic Kirchhoff equations (Eqs.~(\ref{eq:Feq}) and (\ref{eq:Meq})) and using Eqs.~(\ref{eq:da_perturb}) and (\ref{eq:alpha_omega}), we obtain the set of equilibrium equations for the linear perturbation of a helix in the limit of $KL\ll1$:
\begin{eqnarray}
EI\alpha_1 '' &=& - \frac{AB^{\rm r}B^{\rm a}}{\mu_0}\alpha_1\label{eq:om1_helix}\\
EI\alpha_2 '' &=& \frac{AB^{\rm r}B^{\rm a}}{\mu_0}\cos\psi\left(\alpha_2\cos\psi-\alpha_3\sin\psi\right)\label{eq:om2_helix}\\
GJ\alpha_3 '' &=&-\frac{AB^{\rm r}B^{\rm a}}{\mu_0}\sin\psi\left(\alpha_2\cos\psi-\alpha_3\sin\psi\right)\label{eq:om3_helix},
\end{eqnarray}
with boundary conditions that are the same as those given in Sec.~\ref{sec:Twist}: $\alpha_a(0) = \alpha_a '(L)= 0$. 
Equations~(\ref{eq:om1_helix})-(\ref{eq:om3_helix}) are the linearized moment balance equations of Eq.~(\ref{eq:Meq}).

Assuming the functional form $\alpha_a\propto\sin(ks)$ with $k = \pi/2L$ (that satisfies the boundary conditions $\alpha_a(0) = \alpha_a '(L)= 0$), the critical magnetic field $B^{a*}$ satisfying $\alpha_a\ne0$ can be obtained from Eqs.~(\ref{eq:om2_helix}) and (\ref{eq:om3_helix}) as
\begin{eqnarray}
B^{a*} = \left(\frac{\pi}{2L}\right)^2\frac{\mu_0}{AB^{\rm r}}\frac{EI}{\cos^2\psi + (EI/GJ)\sin^2\psi}\label{eq:Bstar_helix}.
\end{eqnarray}
Equation~(\ref{eq:Bstar_helix}) captures the previous results on planar bending in Eq.~(\ref{eq:Bbend}) in Sec.~\ref{sec:Reduc2D}, as well as the twist instabilities in Eq.~(\ref{eq:Btwist}), which we investigated in Sec.~\ref{sec:Twist}. Indeed, by taking the limit $\psi\to 0$, (i.e., rod clamped vertically and magnetized along its tangent), we get $B^{a*}\to B^{a*}_{\rm bend}$. By contrast, in the limit of $\psi\to\pi/2$, (i.e., rod clamped horizontally but magnetized as in Sec.~\ref{sec:Twist}), we recover $B^{a*}\to B^{a*}_{\rm twist}$.

We performed numerical simulations corresponding to the buckling of a hard magnetic helix under the constant magnetic field. When the applied field is sufficiently small $B^{\rm a} < B^{a*}$, the rod remains helical, but the helix buckles when $B^{\rm a} > B^{a*}$. In Figs.~\ref{fig:HelixBuckling} (b) and (c), we show representative snapshots of the numerical simulations for $B^{\rm a} < B^{a*}$ and $B^{\rm a} > B^{a*}$, respectively.  
Based on Eq.~(\ref{eq:Bstar_helix}), we summarize experimental and numerical results in the rescaled phase diagram shown in Fig.~\ref{fig:HelixBuckling} (d). 
In the simulations, we remark that the rod remains helical if the displacement of the free-end is less than $10^{-3}\ell_0$, where $\ell_0$ is the natural length of the spring in the simulation (unit length of the simulation). 
The phase boundary between the undeformed (helical) and buckled states is in an excellent agreement between experiments and simulations. 
Furthermore, the analytic prediction (Eq.~(\ref{eq:Bstar_helix})) correctly predicts both experimental and numerical results. 
It should be noted that not only Eq.~(\ref{eq:Bstar_helix}) predicts the phase boundary for $L/R\ll1$, but also that for $L/R = O(1)$, without any adjustable parameters. 
A more rigorous analysis will be able to derive the full analytic prediction of the phase boundary, wheres it is beyond the scope of the current paper.

In this subsection, we have confirmed that the elasto-magnetic torque $\bm{q}_{\rm mag}$  Eq.~(\ref{eq:q_def}) derived in Sec.~\ref{sec:reduction} correctly predicts the experimental results for a naturally curved and twisted rod. Combining the results from Sec.~\ref{sec:Twist} and those in this subsection, we validated the magnetic Kirchhoff equations (Eqs.~(\ref{eq:Feq}) and (\ref{eq:Meq})) in 3D deformation under the constant field. In the next subsection, we consider the case of loading in a constant gradient magnetic field to validate the elasto-magnetic force $\bm{p}_{\rm mag}$ (Eq.~(\ref{eq:p_def})) to fully validate our theoretical framework.

\subsection{Deformation of a magnetic helix under constant gradient field}
\label{sec:Helix}

In this subsection, we will validate the framework for magnetic helices under the \textit{constant gradient} field (${\bm p}_{\rm mag}, {\bm q}_{\rm mag} \ne 0$). 
For this third and final test configuration, the magnetic helix is clamped at $s = 0$, while setting the other end ($s=L$) to be free.
As we increase the gradient of the field $b$ (or equivalently, the rescaled gradient $\lambda_m = L/\ell_m$, see Eq.~(\ref{eq:magnetobending})), the magnetic helix stretches. 
We will present results on the displacement of the free-end as a function of $\lambda_m$ and find excellent agreement between experiments and simulations.
In this subsection, we study the magnetic helix of a pitch shorter than that in Sec.~\ref{sec:HelixBuckling}. Hence, the helix interacts with itself non-locally via long-range interaction. 
Although long-range interactions are not considered in our theory, we can predict the stretching of the helix as long as the pitch becomes larger upon increasing $\lambda_m$.

\begin{figure}[!h]
    \centering
    \includegraphics[width=1.0\textwidth]{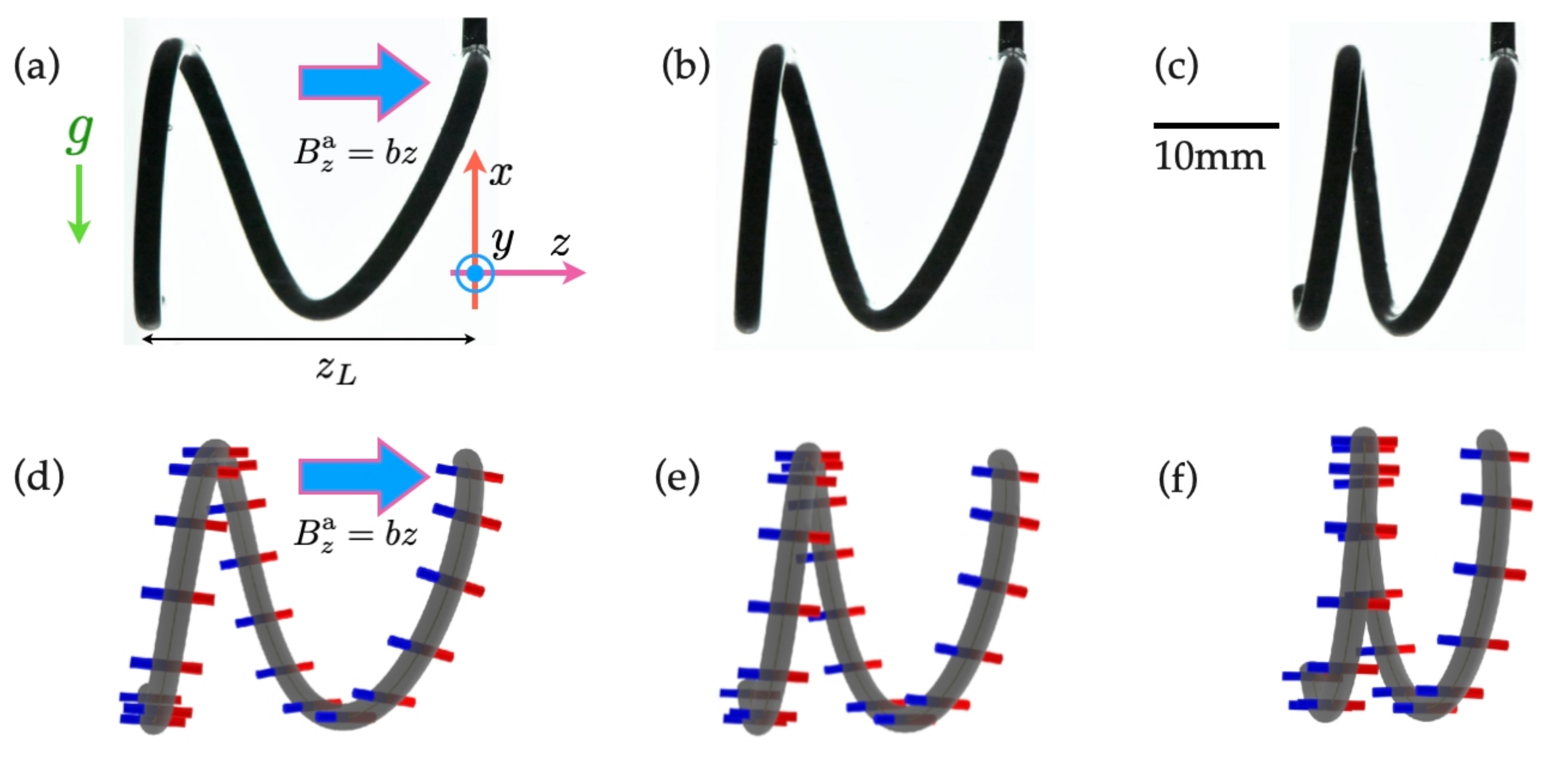}
    \caption{Deformation of a hard magnetic helix. Snapshots of (a)-(c) experiments and (d)-(f) simulations with ${B}^{\rm r} = 13~{\rm mT}$, $R = 10~{\rm mm}$ and $L = 103~{\rm mm}$. The corresponding values of rescaled gradient are (a) (d) $\lambda_m = 3.0$, (b) (e) $\lambda_m = 2.5$, (c) (f) $\lambda_m = 2.0$.}
    \label{fig:Helix1}
\end{figure}

As an illustrative configuration, the magnetic helix is placed with its central axis is aligned with the coil axis, and the clamped-end is located on the $z = 0$ plane (see Fig.~\ref{fig:Helix1} for a photograph of the setup), within the error of $\pm1$~mm.
The magnetic helix has geometries of $(L,R) = (10, 103)$ ${\rm mm}$ and pitch angle $\psi = 1.51~{\rm rad}$.
We suspend the magnetic helix in a glycerol bath, horizontally.
The gradient of the magnetic field $b$ is increased from zero up to $b = 1.06~{\rm T/m}$ (maximum value in our setup), and then $b$ is decreased back to zero.

In the absence of the applied field, $\lambda_m = 0$ (i.e. $b = 0$), the helix is fully contracted due to the long-range dipole-dipole attractive interaction. 
As discussed in Refs.~\cite{Zhao:2019hk,Yan:2020prep} and Sec.~\ref{sec:Theory}, the applied field gradient is analogous to a unidirectional body force of constant magnitude such as gravity. Hence, the helix extends (or contracts) when $\lambda_m$ is increased or decreased, respectively.
We present the photograph of the extended helix with $\lambda_m = 3.0$ in Fig.~\ref{fig:Helix1} (a). The rescaled gradient is then decreased to $\lambda_m = 2.5$ and $\lambda = 2.0$, as shown in Fig.~\ref{fig:Helix1} (b) and (c), respectively. 
We performed the numerical simulations of the magnetic helix under the constant gradient field to compare its deformed configurations with the experimental ones. 
Figures~\ref{fig:Helix1} (d), (e) and (f) are the snapshots of our discrete simulation with $\lambda_m = 3.0, 2.5$ and $2.0$, respectively.
Similar to our experimental observations, the helix contracts as $\lambda_m$ is decreased. 
The predicted shapes in the numerical simulation agree with the configurations in experiments qualitatively.

\begin{figure}[!h]
    \centering
    \includegraphics[width=1.0\textwidth]{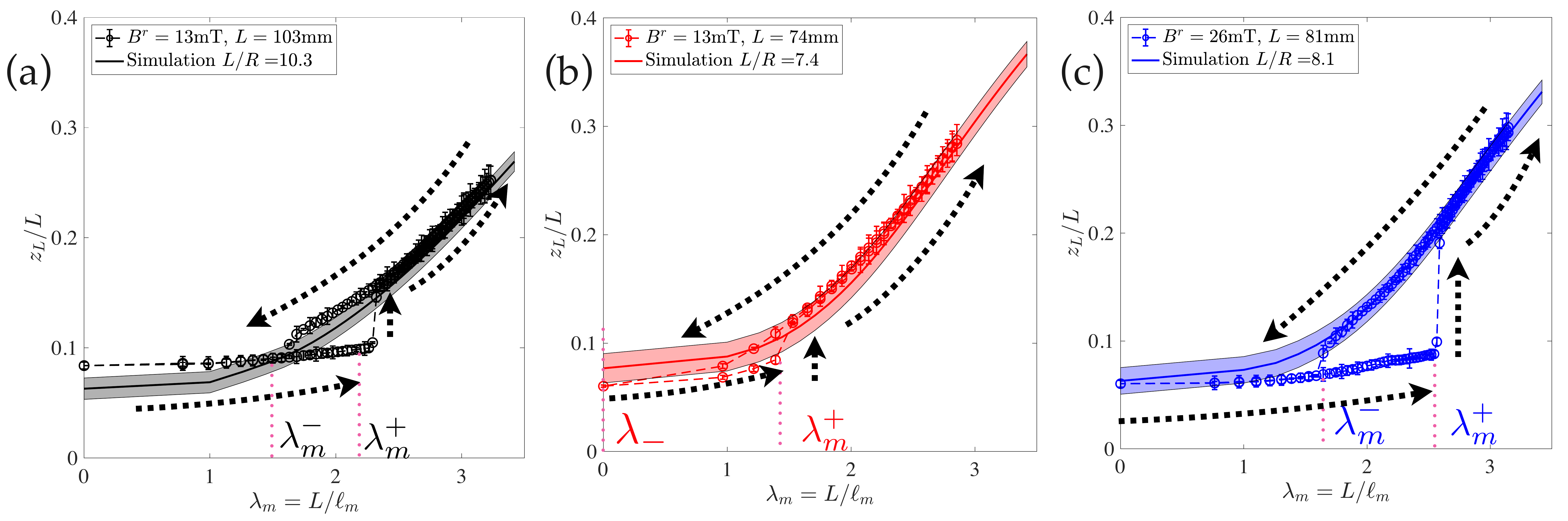}
    \caption{The $z$-position of the free-end as a function of the rescaled gradient $\lambda_m = L/\ell_m$ for different $B^{\rm r}$ and length $L$. The data points and solid lines are experimental and simulation results, respectively. The shaded areas are errors due to the clamped position computed from the simulation. (a) $B^{\rm r} = 13~{\rm mT}$, $L = 103~{\rm mm}$ (b) $B^{\rm r} = 13~{\rm mT}$, $L = 74~{\rm mm}$ (c) $B^{\rm r} = 26~{\rm mT}$, $L = 81~{\rm mm}$. }
    \label{fig:Helix2}
\end{figure}

To quantify the deformation of the helix in the present configuration, we measured the end-to-end distance along $z$ between the clamp and the tip of the helix: $z_L\equiv z(L)-z(0)$ defined in Fig.~\ref{fig:Helix1} (a). 
In Fig.~\ref{fig:Helix2} (a)-(c), we plot $z_L$ as a function of the rescaled applied field gradient $\lambda_m = L/\ell_m$ for different values of $L$ or $B^{\rm r}$, where $\ell_m$ is the magneto-bending length defined in Eq.~(\ref{eq:magnetobending}). 
The data points correspond to the experimental results. 
As an example, we describe the experimental results presented in Fig.~\ref{fig:Helix2} (a). 
In the absence of the applied field $\lambda_m = 0$, the helix is in self-contact due to long-range attractive interactions. Upon increasing $\lambda_m$, the helix loses self-contact when $\lambda_m = \lambda_m ^+$. 
Past this point, $z_L$ increases as the elasto-magnetic force stretches the helix. 
When $\lambda_m$ is decreased, the helix contracts until self-contact at $\lambda_m = \lambda_m ^-$. 
For the sake of convenience, we refer to the branch where the helix is in self-contact \textit{self-contact branch}, and to the branch without self-contact as the \textit{elasto-magnetic branch}.

We find that the values of $\lambda_m ^{\pm}$ depend on $B^{\rm r}$ and $L$.
Indeed, in Figs.~\ref{fig:Helix2} (a) and (c), $\lambda_m ^-\ne0$, while, in Fig.~\ref{fig:Helix2} (b), $\lambda_m ^-=0$. The fact that $\lambda_m ^-=0$ implies that the helical shape (without contact) is stable but it is not a unique stable state due to the long-range interactions. 
The value of $\lambda_m ^{\pm}$ can be used to classify the hysteric behavior, depending on the magnitude of residual flux density $|B^{\rm r}|$, the pitch angle $\psi$, and the rescaled total length $L/R$. 
We will discuss the hysteric behavior in $z_L$ more in detail, in Sec.~\ref{sec:Limitation}. 

Complementing the experimental data, we performed the discrete simulations corresponding to our experimental settings to measure the end-to-end distance $z_L$ as a function of $\lambda_m$. Given that our theory does not include the self long-range interaction, $z_L$ is not hysteretic (the self-contact branch does not appear). 
Still the predictions from our model are in excellent agreement with the experimental results in the elasto-magnetic branches.
In Figs.~\ref{fig:Helix2} (a)-(c), we plot the predictions from the simulations as solid lines.
The width of the shaded region in Figs.~\ref{fig:Helix2} (a)-(c) correspond to the propagation of errors due to the $\pm1$~{\rm mm}, uncertainty of the total length $L$ measurement.
From the excellent agreement between the experimental data and numerical results, we conclude that the magnetic Kirchhoff equations (Eqs.~(\ref{eq:Feq}) and (\ref{eq:Meq})) provide an accurate quantitative prediction for the deformation of the magnetic helix under a constant gradient field, when the self long-range interaction is not dominant.

In this section, we combined the discrete simulations with experiments, to validate the magnetic Kirchhoff equations (Eqs.~(\ref{eq:Feq}) and (\ref{eq:Meq})) derived in Sec.~\ref{sec:vw}, for specific cases where the long-range interaction can be neglected compared with the elasto-magnetic forces. 
We showed that the numerical simulations can predict the behavior of the elasto-magnetic branch observed experimentally. 
We have validated the elasto-magnetic torque $\bm{q}_{\rm mag}$ and force $\bm{p}_{\rm mag}$ in a naturally straight and curved rod, thereby allowing us to validate Eqs.~(\ref{eq:Feq}) and (\ref{eq:Meq}) in 3D geometrically nonlinear deformations, at least when the long-range interaction is negligible.
Next, in Sec.~\ref{sec:Limitation}, we perform a more systematic experimental study of when and how the long-range interaction affects the deformation of a magnetic helix. 

\section{Limitations of the theory}
\label{sec:Limitation}
In this section, we systematically quantify the limitation of the theory through precision experiments alone, because the self long-range interactions is not included in Eqs.~(\ref{eq:Feq}) and (\ref{eq:Meq}); doing so is a challenging endeavour that goes beyond the scope of the present study. We classify the hysteretic behavior reported in Fig.~\ref{fig:Helix2} based on the value of $\lambda_m ^{\pm}$; the critical field strength at which $z_L$ jumps discontinuously. We will discuss the classification by considering the structural instability of the magnetic helix in the absence of the applied field, that is $\lambda_m = 0$.  

The long-range interactions in the magnetic helix depend on the combined effects of the magnitude of the magnetization $|\bm{\mathcal{M}}|\propto|\bm{B}^{\rm r}|$ and the geometry of the helix  (e.g., the pitch angle $\psi$ and the rescaled total length $L/(2\pi R)$). 
The magnitude of the magnetization $|\bm{\mathcal{M}}|$ is programmed at the fabrication stage by the concentration ratio of the NdPrFeB particles $c$. 
When $c$ is small, the long-range interaction will be suppressed (and vice versa). 
On the other hand, the dependence of $\psi$ and $L/(2\pi R)$ is highly nontrivial, as we discuss below.

The constant gradient field is applied to the magnetic helix as in Sec.~\ref{sec:Helix}. 
The helix is extended and contracted under the constant gradient magnetic field, for several pitch angles (in the natural configuration). 
We introduce the normalized pitch angle as $\tilde{\psi}\equiv\psi/(\pi/2)$ such that $\tilde{\psi} = 1$ corresponds to the planar circle with zero twist (experimentally non-realizable due to the finite diameter of the rod) and $\tilde{\psi} = 0$ corresponds to a straight rod. As $\tilde{\psi}$ decreases, the pitch of the helix increases.  
Practically, $\tilde{\psi}_{\rm max}=0.96$ is the largest value of $\tilde{\psi}$ realizable in our experimental conditions with $(d,R) = (2,10)$ ${\rm mm}$. 
Our rods are fabricated systematically through the procedure in Sec.~\ref{sec:fab} with different pitch angles in the range $0.80\leq\tilde{\psi}\leq0.95$. 
We clamp the magnetized helix with $B^{\rm r} = 26~{\rm mT}$ at $z(0) = 0$, while $s = L$ is set to be free. Through the cyclic protocols consisting of increasing and decreasing the flux density, the values of $\lambda_m$ characterizing the hysteresis $\lambda_m ^{\pm}$ are identified (see Fig.~\ref{fig:Helix2} for the definition of $\lambda_m ^{\pm}$).

\begin{figure}[!t]
    \centering
    \includegraphics[width=0.7\textwidth]{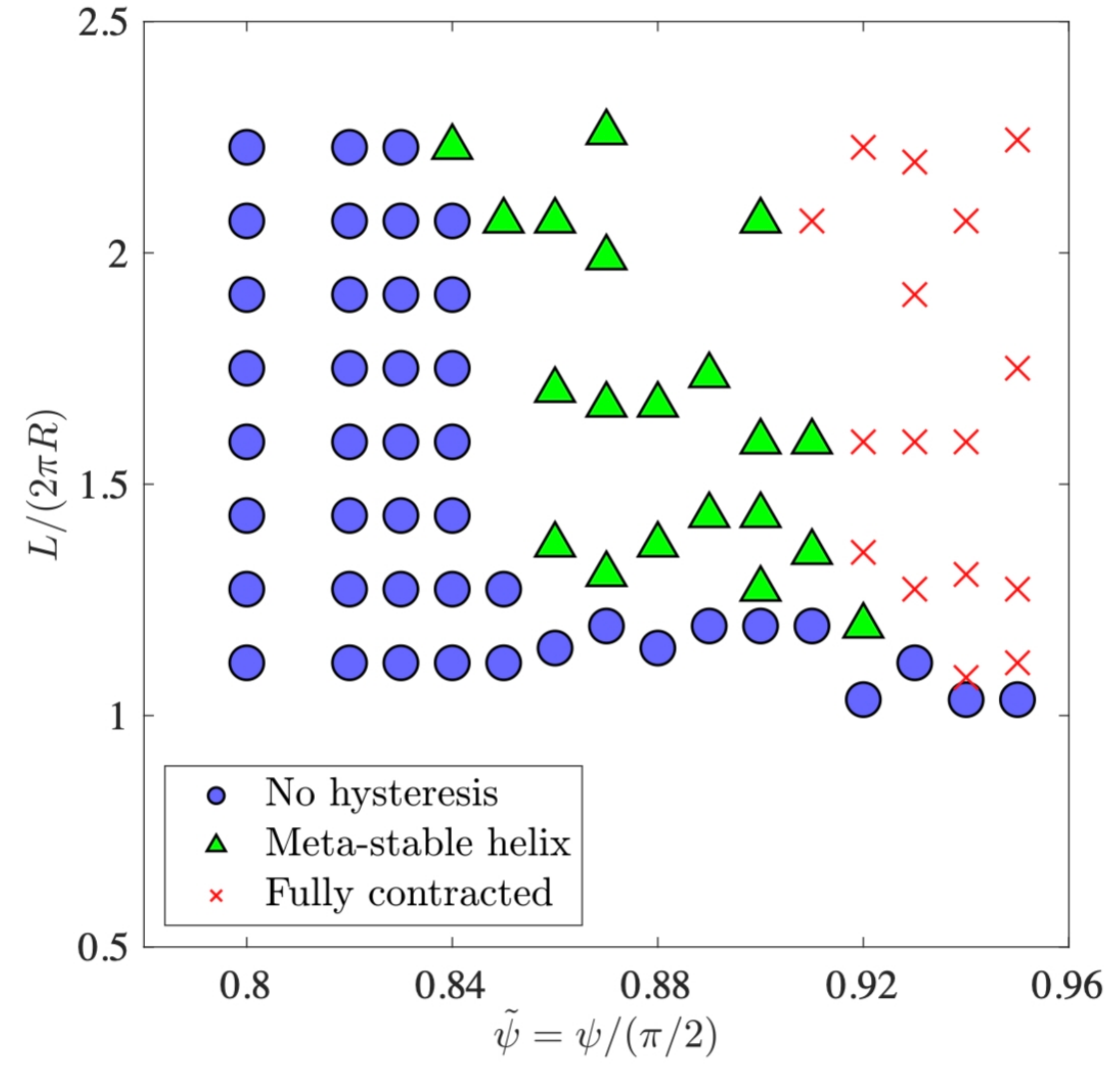}
    \caption{Classification of the hysteretic behavior in the stretching and contraction of a magnetic helix. The phase diagram of hysteresis is obtained from experiments with $B^{\rm r} = 26{\rm mT}$ and $R = 10{\rm mm}$; $\bigcirc$: No hysteresis ($\lambda_m^+,\lambda_m ^- = 0$), $\triangle$: Meta-stable helix ($\lambda_m^+\ne 0,~\lambda_m ^- = 0$) and $\times$: Fully contracted helix ($\lambda_m^+,\lambda_m ^- \ne 0$).  
    }
    \label{fig:Hysteresis}
\end{figure}

In Fig.~\ref{fig:Hysteresis}, we present the experimental phase diagram in the $\tilde{\psi}$-$L/2\pi R$ parameter space to classify the response against $\lambda_m$, finding that there are three different possible responses (i)-(iii), depending on the geometry of the magnetic helix. 
(i)~When $L/2\pi R\simeq 1$ (length of the helix is nearly a single turn) or  $\tilde{\psi}$ is small, (i.e. when the pitch $p = 2\pi R/\tan\psi$ is long enough), no hysteresis is observed (circles in Fig.~\ref{fig:Hysteresis}). 
When we plot $z_L$ as a function of $\lambda_m$, $z_L$ changes continuously and reversibly for any $\lambda_m$.  The tip of the helix follows the same curve upon increasing or decreasing $\lambda_m$.
In this case, we have $\lambda_m ^+ = \lambda_m ^- = 0$ (Fig.~\ref{fig:Hysteresis}).
For higher values of $\psi$, we observe hysteresis, which is divided into two further categories; $\lambda_m^+\ne 0,~\lambda_m ^- = 0$ or $\lambda_m^+,\lambda_m ^- \ne 0$, which are shown as triangles and cross marks in Fig.~\ref{fig:Hysteresis}, respectively.  
(ii)~In the case of $\lambda_m^+\ne 0,~\lambda_m ^- = 0$, $z_L$ changes discontinuously from the contracted helix to the stretched helix at $\lambda_m = \lambda_m ^+$, while $z_L$ decreases smoothly as $\lambda_m\to0$, i.e. $\lambda_m ^- = 0$ (see Fig.~\ref{fig:Helix2} (b) for the corresponding behavior). 
(iii)~By contrast, in the case of $\lambda_m^+,\lambda_m ^- \ne 0$, $z_L$ jumps at both $\lambda_m ^+$ and $\lambda_m ^-$. Thus, if we plot $z_L$ as a function of $\lambda_m$, we obtain similar curves as those in Figs.~\ref{fig:Helix2} (a) and (c).

The classification of the hysteretic responses in the $z_L$-$\lambda_m$ plots (i)-(iii) can be understood by considering the configuration at $\lambda_m = 0$. 
In the following, we discuss the stability of the uniform helix at $\lambda_m = 0$ against mechanical perturbations. 
(i)~When $\lambda_m^+= \lambda_m ^-= 0$ (no hysteresis), the uniform helical shape corresponds to a unique energy minimum. The reference configuration is mono-stable upon any mechanical perturbations, because self-interactions have minimal effects. 
(ii)~For $\lambda_m^+ \ne 0$ and $ \lambda_m ^-= 0$, the uniform helix is stable but it is not a unique minimum. In this case, the helix can contracts to the (stable) state with the normalized pitch angle of $\tilde{\psi} = \tilde{\psi}_{\rm max}$ if perturbed; we call this state {\it meta-stable helix}. 
Lastly, (iii)~for $\lambda_m^+,\lambda_m^- \ne 0$, the uniform helix is no longer stable. Indeed, the contracted state is the unique energy minimum. 
In this third case, the helix remains fully-contracted against any mechanical perturbation.

The nature of the hysteretic behavior in magnetic helices depends on the concentration ratio of NdPrFeB particles $c$ and the geometry of the helix as described by $\psi$ and $L/(2\pi R)$.  
We classified the deformation due to long-range self-interactions based on the value of $\lambda_m$ at which $z_L(\lambda_m)$ changes discontinuously. 
The hysteretic behavior in the magnetic helix originates from the bi-stability due to the long-range interactions. In the previous sections, we had validated the magnetic Kirchhoff rod equations when the long-range interaction can be neglected. At the same time, we clarified the limitation of our theory systematically. The hysteretic behavior, which our theory cannot capture, is observed only when the pitch angle $\tilde{\psi}$ is large and when the rod is long $L/(2\pi R) \gtrsim 1$. 
We can conclude that the magnetic Kirchhoff rod equations (Eqs.~(\ref{eq:Feq}) and (\ref{eq:Meq})) are valid as long as $\psi$, $L/(2\pi R)$, or $c$ remains small.

\section{Conclusion}
\label{sec:Concl}
The goal of the present study was to develop a Kirchhoff-like theory for hard magnetic rods based on dimensional reduction. 
Our theoretical framework, rooted in 3D elasticity, yielded a reduced (centerline-based) description of the magnetic rod~(Sec.~\ref{sec:Theory}). The set of governing equations contain elasto-magnetic forces and torques (Eqs.~(\ref{eq:p_def}) and (\ref{eq:q_def})), in addition to the purely elastic ones for a (non-magnetic) rod.
We validated the magnetic Kirchhoff rod equations (Eqs.~(\ref{eq:Feq}) and (\ref{eq:Meq})) through a set of precision experiments. 
Our theoretical results supplemented by the discrete simulation are in excellent agreement with the experimental results.
Moreover, the magnetic Kirchhoff rod equations reproduce the previous results on the planar deformation of hard magnetic beams or elastica~\cite{Lum:2016fc,Zhao:2019hk,Yan:2020prep,Ciambella:2020je}. 
We confirmed that the force and moment balance equations in our theory for hard magnetic rods reduce to those of  Refs.~\cite{Lum:2016fc,Zhao:2019hk,Yan:2020prep,Ciambella:2020je} for beams and elastica.

To validate our theory for the 3D deformation of magnetic rods, we performed three different sets of experiments; (i)~a straight magnetic rod under constant external magnetic field~(Sec.~\ref{sec:Twist}), (ii)~a helical rod under constant field~(Sec.~\ref{sec:HelixBuckling}), and (iii)~a helical rod under the constant gradient field~(Sec.~\ref{sec:Helix}), with the clamped-free boundary conditions. 
(i)~In the case of a straight rod under the constant field, we studied the twist instability. When the rod is magnetized along $\hat{\bm{d}}_1$ (perpendicular to the tangent $\hat{\bm{d}}_3$) and we apply an anti-parallel field, the rod twists, while its centerline remains straight. We derived an analytical prediction for the critical applied field for this twist instability $B^{\rm a*}_{\rm twist}$, which is in excellent agreement with experimental results. 
(ii)~We studied the buckling instability of a helical rod under constant external magnetic field. The helix was magnetized along its central axis. When we apply the field to the magnetic helix, it buckles above the critical applied field $B^{\rm a*}$.
We showed that this instability is triggered by a balance between elastic and elasto-magnetic torques. 
The critical applied fields were predicted analytically from a linear stability analysis. 
The simulation results correctly capture the experimental observations.
(iii)~Under a constant gradient field, the helix magnetized along the central axis stretches (or contracts) as we increase (or decrease) the applied field strength.
Due to the long-range self-interaction between magnetic dipoles within the rod, the deformation is hysteretic upon extension and contraction.

Our framework correctly predicts the deformation of the magnetic helix when its pitch is large enough or when the long-range self-interactions are negligible. 
The hysteretic behavior was studied systematically through experiments by controlling the pitch angle $\psi$ of the helix. 
We revealed that there are three types of behavior; no hysteresis state, meta-stable, and fully contracted helices, depending on the total length $L/R$ and the pitch angle $\psi$.

Although our theory is in excellent agreement with the experimental results when the elasto-magnetic force and torque are dominant, the non-local interactions between the magnetization vectors $\bm{\mathcal{M}}$ are necessary to describe the hysteretic behavior. 
In Refs.~\cite{Hall:2013gm,Vella:2014ix,Schonke:2017if}, the continuum mechanics of {\it magnetic chains} has been studied by considering the self-interactions between magnetic beads. In the future, combining the continuum mechanics of the magnetic dipole interactions and our magnetic Kirchhoff rod equations, the hysteretic behavior of the magnetic helix could be studied in detail theoretically. Extending the magnetic Kirchhoff equations to include the dipole-dipole interactions is one of the interesting directions of future works.
Also, extending our theoretical framework toward dynamics of hard magnetic rods is also another exciting opportunity for future work. 
The dynamics could be simulated by considering the (external) viscous force and torque of a surrounding medium in Eqs.~(\ref{eq:Feq}) and (\ref{eq:Meq}). 

The framework established in this paper would be valuable to simulate the large deformation of hard magnetic rods used in micro or soft-robotics~\cite{Diller:2014fd,Tsumori:2015ke,Huang:2016ee,Ciambella:2020je} or haptic devices~\cite{Pece:2017ib}. In the future, novel functional devices of complex geometries could be developed by combining the simple building blocks studied in this paper.


\section*{Acknowledgement}
T.G.S. was supported by Grants-in-Aid for Japan Society for the Promotion of Science Overseas Research Fellowship 2019-60059 and MEXT KAKENHI 18K13519.

\appendix

\section{Derivation of Equation~(\ref{eq:alpha_omega})}
\label{sec:App1}
In this appendix, we provide the detailed derivation of Eq.~(\ref{eq:alpha_omega});
\begin{eqnarray}
\alpha_a' = \Omega_a\,.
\end{eqnarray}
We recall that our rod is naturally curved and twisted: ${\bm \Omega}^{(0)} = \Omega_a ^{(0)}\hat{\bm{d}}_a ^{(0)}$. We expand the Cosserat frame basis in Eq.~(\ref{eq:da_perturb}) around the reference configuration $\hat{\bm{d}}_a ^{(0)}$. 
We will determine the relationship between the first-order perturbation of the components of the Darboux vector $\Omega_a ^{(1)}$ defined as
\begin{eqnarray}
{\Omega}_a = {\Omega}^{(0)} _a + {\Omega}^{(1)} _a + \cdots\,,\label{eq:OmegaPerturb}
\end{eqnarray}
and the first order perturbation of $\hat{\bm{d}}_a$, $\alpha_a$. 
Note that $\Omega_a ^{(1)}$ and $\alpha_a$ are infinitesimal quantities of the same order $\Omega_a ^{(1)}\sim\alpha_a$. 
We substitute Eqs.~(\ref{eq:da_perturb})
\begin{eqnarray}
\hat{\bm{d}}_a = (\delta_{ab} + \varepsilon_{abc}\alpha_c)\hat{\bm{d}}_b ^{(0)}\,,
\end{eqnarray}
and (\ref{eq:OmegaPerturb}) into Eq.~(\ref{eq:kinematic}); $\hat{\bm{d}}_a' = \bm{\Omega}\times\hat{\bm{d}}_a$, to derive Eq.~(\ref{eq:alpha_omega}).
The final result is consistent with that in Ref.~\cite{Goriely:1997hw} (which is based on matrix representation), while our derivation is based on a tensor representation. 

First, the kinematic equation Eq.~(\ref{eq:kinematic}) is rewritten using the Eddington epsilon as
\begin{eqnarray}
\hat{\bm{d}}_a ' = - \varepsilon_{abc}\Omega_b\hat{\bm{d}}_c,
\label{eq:kinematic_tensor}
\end{eqnarray}
which also holds for the base solution as $\hat{\bm{d}}_a ^{(0)'} = - \varepsilon_{abc}\Omega_b ^{(0)}\hat{\bm{d}}_c ^{(0)}$.
The left-hand side of Eq.~(\ref{eq:kinematic_tensor}) is computed as
\begin{eqnarray}
\hat{\bm{d}}_a ' 
&=& \{ (\delta_{ab} + \varepsilon_{abc}\alpha_c)\hat{\bm{d}}_b ^{(0)}\}'\nonumber\\
&=& - \varepsilon_{abc}\Omega_b ^{(0)}\hat{\bm{d}}_c ^{(0)} + \varepsilon_{abc}\alpha_c ' \hat{\bm{d}}_b ^{(0)} - \varepsilon_{abc}\varepsilon_{bde}\alpha_c \Omega_d ^{(0)}\hat{\bm{d}}_e ^{(0)}\nonumber\\
&=& - \varepsilon_{abc}\Omega_b ^{(0)}\hat{\bm{d}}_c ^{(0)} + \varepsilon_{abc}\alpha_c ' \hat{\bm{d}}_b ^{(0)} - (\delta_{cd}\delta_{ae} - \delta_{ce}\delta_{ad})\alpha_c\Omega_d ^{(0)}\hat{\bm{d}}_e ^{(0)}\nonumber\\
&=& - \varepsilon_{abc}\Omega_b ^{(0)}\hat{\bm{d}}_c ^{(0)} + \varepsilon_{abc}\alpha_c ' \hat{\bm{d}}_b ^{(0)} - (\alpha_c\Omega_c ^{(0)})\hat{\bm{d}}_a ^{(0)} + \Omega_a ^{(0)} (\alpha_c\hat{\bm{d}}_c ^{(0)}).\label{eq:PerturbLHS}
\end{eqnarray}
We expand the right-hand side of Eq.~(\ref{eq:kinematic_tensor}) as
\begin{eqnarray}
- \varepsilon_{abc}\Omega_b\hat{\bm{d}}_c 
&=& - \varepsilon_{abc}({\Omega}^{(0)} _b + {\Omega}^{(1)} _b)(\delta_{cd} + \varepsilon_{cde}\alpha_e)\hat{\bm{d}}_d ^{(0)}\nonumber\\
&=& - \varepsilon_{abc}\Omega_b ^{(0)}\hat{\bm{d}}_c ^{(0)} - \varepsilon_{abc}\Omega_b ^{(1)}\hat{\bm{d}}_c ^{(0)} - \varepsilon_{abc}\varepsilon_{cde}\alpha_e {\Omega}^{(0)} _b\hat{\bm{d}}_d ^{(0)}\nonumber\\
&=& - \varepsilon_{abc}\Omega_b ^{(0)}\hat{\bm{d}}_c ^{(0)} - \varepsilon_{abc}\Omega_b ^{(1)}\hat{\bm{d}}_c ^{(0)} - (\delta_{ad}\delta_{be} - \delta_{ae}\delta_{bd})\alpha_e {\Omega}^{(0)} _b\hat{\bm{d}}_d ^{(0)}\nonumber\\
&=& - \varepsilon_{abc}\Omega_b ^{(0)}\hat{\bm{d}}_c ^{(0)} - \varepsilon_{abc}\Omega_b ^{(1)}\hat{\bm{d}}_c ^{(0)} - (\alpha_c\Omega_c ^{(0)})\hat{\bm{d}}_a ^{(0)}  +  \alpha_a(\Omega_c ^{(0)}\hat{\bm{d}}_c ^{(0)}).
\label{eq:PerturbRHS}
\end{eqnarray}
Note that, in Eqs.~(\ref{eq:PerturbLHS}) and (\ref{eq:PerturbRHS}), we dropped higher order terms (e.g.~the terms proportional to $\alpha_a\Omega_b ^{(1)}$).

Substituting Eqs.~(\ref{eq:PerturbLHS}) and (\ref{eq:PerturbRHS}) into Eq.~(\ref{eq:kinematic_tensor}), we obtain the tensorial relation
\begin{eqnarray}
\varepsilon_{abc}\alpha_c ' \hat{\bm{d}}_b ^{(0)} 
= -\varepsilon_{abc}\Omega_b ^{(1)}\hat{\bm{d}}_c ^{(0)} 
+  \alpha_a(\Omega_c ^{(0)}\hat{\bm{d}}_c ^{(0)}) - \Omega_a ^{(0)} (\alpha_c\hat{\bm{d}}_c ^{(0)}).
\label{eq:App1Comp}
\end{eqnarray}
Next, by taking the inner product between Eq.~(\ref{eq:App1Comp}) and $\hat{\bm{d}}_d ^{(0)}$, and by arranging the indices, we arrive at
\begin{eqnarray}
\varepsilon_{abc}\alpha_c ' = \varepsilon_{abc}\Omega_c ^{(1)} + \left(\alpha_a\Omega_b ^{(0)} - \Omega_a ^{(0)}\alpha_b\right).
\label{eq:App1Final}
\end{eqnarray}
The presence of the term inside the parenthesis in the right-hand side of Eq.~(\ref{eq:App1Final}) encodes the geometric non-linearity intrinsic to a naturally curved rod. When the rod is naturally straight, $\Omega_a ^{(0)} = 0$, we recover Eq.~(\ref{eq:alpha_omega}). 

\section{Derivation of Equations~(\ref{eq:d1_0})-(\ref{eq:d3_0})}
\label{sec:App2}
In this appendix, we derive the Cosserat frame basis for the uniform helix Eqs.~(\ref{eq:d1_0})-(\ref{eq:d3_0}). To do so, we will integrate the kinematic relation Eq.~(\ref{eq:kinematic}) with ${\bm \Omega} = \kappa\hat{\bm{d}}_2 + \tau\hat{\bm{d}}_3$. The goal is to express $\bm{r}(s)$ and $\hat{\bm{d}}_a~(a=1,2,3)$ in terms of the Cartesian basis $\hat{\bm{e}}_i$. 
From the kinematic relation, we get 
\begin{eqnarray}
\hat{\bm{d}}_1' &=& -\kappa\hat{\bm{d}}_3 + \tau\hat{\bm{d}}_2\label{eq:FS1},\\
\hat{\bm{d}}_2' &=& -\tau\hat{\bm{d}}_1,\label{eq:FS2}\\
\hat{\bm{d}}_3' &=& \kappa\hat{\bm{d}}_1,\label{eq:FS3}
\end{eqnarray}
which are the same as the Frenet-Serret equations~\cite{Audoly:2010Book}.
Differentiating Eq.~(\ref{eq:FS1}) with respect to $s$, and with the help of Eqs.~(\ref{eq:FS2}) and (\ref{eq:FS3}), we find the ODE only for $\hat{\bm{d}}_1$;
\begin{eqnarray}
\hat{\bm{d}}_1'' = - K^2\hat{\bm{d}}_1\,.
\label{eq:d1_ODE2}
\end{eqnarray}
Here, we introduced the normalized curvature $K$ as $K=\sqrt{\kappa^2 + \tau^2}$, or equivalently
\begin{eqnarray}
\kappa = K \sin\psi,~~\tau = K\cos\psi.
\label{eq:Kvskappatau}
\end{eqnarray}

After integrating Eq.(~\ref{eq:d1_ODE2}) twice, we compute $\hat{\bm{d}}_1$ as
\begin{eqnarray}
\hat{\bm{d}}_1(s) = \cos(Ks)\hat{\bm{d}}_1(0) + \frac{\tau}{K}\sin(Ks)\hat{\bm{d}}_2(0) - \frac{\kappa}{K}\sin(Ks)\hat{\bm{d}}_3(0).
\end{eqnarray}
Following a similar procedure for $\hat{\bm{d}}_2$ and $\hat{\bm{d}}_3$, we obtain
\begin{eqnarray}
\hat{\bm{d}}_2(s)-\hat{\bm{d}}_2(0) &=& - \frac{\tau}{K}\sin(Ks)\hat{\bm{d}}_1(0) - \frac{\tau}{K^2}(1-\cos(Ks))(\tau\hat{\bm{d}}_2(0) - \kappa\hat{\bm{d}}_3(0)),\\
\hat{\bm{d}}_3(s)-\hat{\bm{d}}_3(0) &=&  \frac{\kappa}{K}\sin(Ks)\hat{\bm{d}}_1(0) + \frac{\kappa}{K^2}(1-\cos(Ks))(\tau\hat{\bm{d}}_2(0) - \kappa\hat{\bm{d}}_3(0)).
\end{eqnarray}
By integrating $\boldsymbol{r}' = \hat{\bm{d}}_3$, the centerline position is obtained as
\begin{eqnarray}
\boldsymbol{r}(s) -\boldsymbol{r}(0) = \frac{\kappa}{K^2}(1-\cos(Ks))\hat{\bm{d}}_1(0) - \frac{\kappa}{K^3}\sin(Ks)(\tau\hat{\bm{d}}_2(0) - \kappa\hat{\bm{d}}_3(0)) + \frac{\tau s}{K^2}(\kappa\hat{\bm{d}}_2(0) + \tau\hat{\bm{d}}_3(0)),\nonumber\\ 
\label{eq:helix_v1}
\end{eqnarray}
where the unit vector of the central axis of the helix $\hat{N}$ is
\begin{eqnarray}
\hat{N}\equiv \frac{\kappa}{K}\hat{\bm{d}}_2(0) + \frac{\tau}{K}\hat{\bm{d}}_3(0).
\end{eqnarray}

Finally, let us rewrite Eq.~(\ref{eq:helix_v1}) in a more compact form by choosing $\hat{\bm{d}}_a(0)$ and $\boldsymbol{r}(0)$ such that the central axis lies along $-\hat{\bm{e}}_z$; i.e., we select $\hat{N} = -\hat{\bm{e}}_z$, $\hat{\bm{d}}_1(0) = \hat{\bm{e}}_x$, and $\hat{\bm{e}}_y = \hat{\bm{e}}_z\times\hat{\bm{e}}_x = -(\tau\hat{\bm{d}}_2(0) - \kappa\hat{\bm{d}}_3(0))/K$. Then, using Eq.~(\ref{eq:Kvskappatau}), the centerline position is calculated as
\begin{eqnarray}
\boldsymbol{r}(s) = -\frac{\sin\psi}{K}\cos(Ks)\hat{\bm{e}}_x + \frac{\sin\psi}{K}\sin(Ks)\hat{\bm{e}}_y - s\cos\psi\hat{\bm{e}}_z,
\end{eqnarray}
which implies
\begin{eqnarray}
R = \frac{\sin\psi}{K}.
\end{eqnarray}
The normalized curvature $K$ is now rewritten as a function of $\kappa$ and $\psi$ only; $K = \kappa\sqrt{1 + \tan^{-2}\psi} = \kappa/\sin\psi$, i.e. $\kappa = \sin^2\psi/R$ and $\tau = \sin\psi\cos\psi/R = \sin2\psi/2R$. The choice of $\hat{N}$ allows us to obtain the Cosserat frame basis vectors as
\begin{eqnarray}
\hat{\bm{d}}_1(s) &=& \cos(Ks)\hat{\bm{e}}_x - \sin(Ks)\hat{\bm{e}}_y,\\
\hat{\bm{d}}_2(s) &=& -\cos\psi\left(\sin(Ks)\hat{\bm{e}}_x + \cos(Ks)\hat{\bm{e}}_y\right) - \sin\psi\hat{\bm{e}}_z,\\
\hat{\bm{d}}_3(s) &=& \sin\psi\left(\sin(Ks)\hat{\bm{e}}_x + \cos(Ks)\hat{\bm{e}}_y\right) - \cos\psi\hat{\bm{e}}_z,
\end{eqnarray}
thereby reproducing Eqs.~(\ref{eq:d1_0})-(\ref{eq:d3_0}) that we set out to derive.



\bibliographystyle{model1-num-names}

\bibliography{references.bib}







\end{document}